\documentclass[format=acmsmall, review=false, screen=true]{acmart}

\usepackage{booktabs} 
\usepackage{todonotes}
\usepackage{graphicx}
\usepackage{subfigure}
\usepackage{enumerate}
\usepackage{courier}
\usepackage{multirow}
\usepackage{color}
\usepackage{amssymb}
\usepackage{amsmath}
\usepackage{array}
\usepackage{xcolor}
\usepackage{xcolor, colortbl}
\usepackage{tikz}
\usepackage[group-separator={,}]{siunitx}
\usepackage{enumitem}
\usepackage{hhline}

\usepackage[ruled]{algorithm2e} 

\SetAlFnt{\small}
\SetAlCapFnt{\small}
\SetAlCapNameFnt{\small}
\SetAlCapHSkip{0pt}
\IncMargin{-\parindent}

\acmJournal{TIST}

\setcopyright{acmlicensed}

\acmDOI{0000001.0000001}

\begin{document}
\title[Predicting Academic Performance for College Students: A Campus Behavior Perspective]{Predicting Academic Performance for College Students: A Campus Behavior Perspective}

\author{Huaxiu Yao}
\affiliation{%
  \institution{University of Electronic Science and Technology of China}
 \city{Chengdu}
 \country{China}}
\email{yhx662012@gmail.com}
\author{Defu Lian}\authornote{corresponding author}
\affiliation{%
  \institution{University of Electronic Science and Technology of China}
  \city{Chengdu}
  \country{China}
}
\email{dove.ustc@gmail.com}
\author{Yi Cao}
\affiliation{%
 \institution{University of Electronic Science and Technology of China}
 \city{Chengdu}
 \country{China}}
\email{caoyi318@gmail.com}

\author{Yifan Wu}
\affiliation{%
  \institution{University of Electronic Science and Technology of China}
  \city{Chengdu}
  \country{China}
}
\email{wuyifan9204@qq.com}
\author{Tao Zhou}
\affiliation{%
  \institution{University of Electronic Science and Technology of China}
  \city{Chengdu}
  \country{China}}
\email{zhutou@ustc.edu}

\begin{abstract}
    Detecting abnormal behaviors of students in time and providing personalized intervention and guidance at the early stage is important in educational management. Academic performance prediction is an important building block to enabling this pre-intervention and guidance. Most of the previous studies are based on questionnaire surveys and self-reports, which suffer from small sample size and social desirability bias. In this paper, we collect longitudinal behavioral data from $6,597$ students' smart cards and propose three major types of discriminative behavioral factors, diligence, orderliness, and sleep patterns. Empirical analysis demonstrates these behavioral factors are strongly correlated with academic performance. Furthermore, motivated by social influence theory, we analyze the correlation between each student's academic performance with his/her behaviorally similar students'. Statistical tests indicate this correlation is significant. Based on these factors, we further build a multi-task predictive framework based on a learning-to-rank algorithm for academic performance prediction. This framework captures inter-semester correlation, inter-major correlation and integrates student similarity to predict students' academic performance. The experiments on a large-scale real-world dataset show the effectiveness of our methods for predicting academic performance and the effectiveness of proposed behavioral factors.
\end{abstract}

%
%


%
%


\maketitle

\renewcommand{\shortauthors}{H. Yao et al.}

\section{Introduction}
Education is the foundation of a nation. One important task of educational research is the early prediction of academic performance, which not only helps educators design in-time intervention but also facilitates personalized education. The major challenge is to reveal important factors that affect students' academic performance. It has been demonstrated that physical status~\cite{taras2005obesity,mo1999school}, intelligence quotient~\cite{deary2007intelligence} and even socioeconomic status~\cite{white1982relation} are correlated with academic performance. However, these characteristics are relatively stable over the long run and are difficult to change via educational management. 

Comparatively, more studies are focused on the perspectives of psychology and behavior, partially due to the possibility of intervening on student's mentation and behavior. Extensive experiments about the correlation between the big-five personality traits (Openness, Conscientiousness, Extraversion, Agreeableness, Neuroticism) and academic performance have been reported~\cite{vedel2014big,poropat2009meta,BARRICK:PEPS1}, uncovering conscientiousness as one of the strongest predictors. Behaviors like class attendance~\cite{crede2010class}, lifestyle~\cite{wald2014associations} and sleep habit~\cite{dewald2010influence} are also highly associated with academic performance. However, almost all of these results are obtained from questionnaires or self-reports, which usually suffer from small sample size and social desirability bias, resulting in the difficulty to draw a valid and solid conclusion.

Thanks to the development of information technology, there is a growing trend to augment physical facilities with sensing, computing and communication capabilities in modern universities. These facilities provide an unprecedented opportunity to collect real-time digital records about students' campus activities in an unobtrusive way and to reveal behavioral predictors for academic performance. In this paper, mainly through campus smart card, we collect $6,597$ students' longitudinal behavioral data spanning almost three years. These behaviors include entry-exit of library and dormitory, book borrowing, consumption on campus places (e.g., canteen, supermarket and teaching building). 

From these records, we quantify diligence and orderliness as two kinds of behavioral predictors for academic performance, which is motivated by the strong effect of conscientiousness on academic performance~\cite{dudley2006meta}. The diligence predictors estimate how much time students spend on the study while the orderliness predictors quantify the regularity of students' life and study on campus. Furthermore, according to the previous study about the correlation between students sleep patterns and their academic performance~\cite{trockel2000health,dewald2010influence}, we extract students' sleep patterns from our dataset and regard it as another important predictor. Empirical analysis shows these behavioral predictors are strongly correlated with academic performance (e.g., the averaged Spearman correlation of diligence over semesters achieve as much as $-0.308$).

Motivated by social influence theory, we further analyze the correlation of each student's academic performance with his/her behaviorally similar students'. As suggested in~\cite{yao2017predicting}, student's behavioral similarity is computed based on their co-occurrence at the same location within a short time interval. 
We test the significance of this correlation via t-test and the result indicates that this correlation is significant. Hence, students with similar lifestyles also have close academic performance.

These observations and analysis motivate us to predict students' academic performance for helping education administrators detect undesirable abnormal behaviors in time and implement effective interventions. We aim to tackle the following three challenges: (1) These factors' predictive strength changes across semesters and among different majors, how to model these changes for academic performance prediction; 
(2) The number of students varies from major to major (from 50 to 600), thus the amount of training data in some majors are limited to train a good prediction model; 
(3) It is possible that some students don't have sufficient behavioral data so that behavioral factors could be unreliably measured.

\begin{figure}[h]
	\centering
	\includegraphics[width=0.55\textwidth]{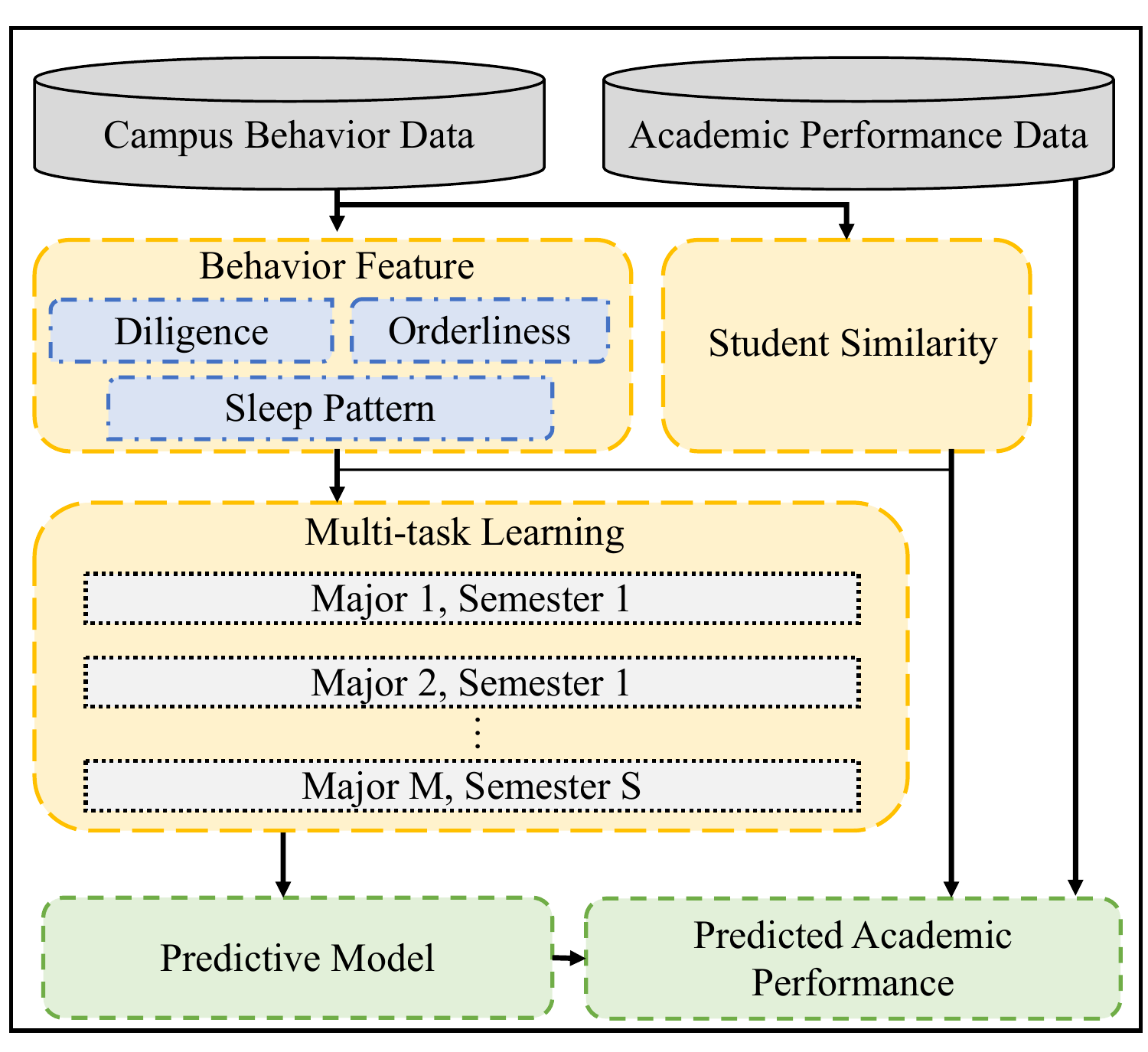}
	\caption{The framework of proposed MTLTR-APP. First, three behavior features are collected based on campus behavior data: diligence, orderliness, and sleep pattern. Then, the value of student similarity is calculated by the campus behavior data. Next, the multi-task learning framework is proposed to predict the score of each student. Finally, by combining the score with behavioral similar students, the final predicted academic performance is presented. We use actual academic prediction data to evaluate our model.}
	\label{fig:framework}
\end{figure}

To address the aforementioned challenges, we propose a novel \textbf{M}ulti-\textbf{T}ask \textbf{L}earning-\textbf{T}o-\textbf{R}ank \textbf{A}cademic \textbf{P}erformance \textbf{P}rediction framework (MTLTR-APP) for predicting students' academic performance. 
Specifically, we impose a sequential smoothing regularization term to model the inter-semester temporal correlation. In addition, though factors' predictive strength for academic performance is varied from major to major, there is still commonality among similar majors (e.g., computer science and electronic engineering). We construct a matrix-factorization-based multi-task model to capture the inter-major correlation for helping the prediction of majors with limited training data. Our framework further incorporates student similarity to help the prediction of students without sufficient behavioral data.
Figure~\ref{fig:framework} illustrates the whole framework of our MTLTR-APP. Specifically, based on campus behavior data, we extract three types of features and calculate student behavioral similarity value. Then we use the proposed multi-task learning model and integrate it with student similarity to predict students' academic performance. 

Finally, we train the proposed predictive algorithm on a large-scale behavioral dataset from a grade of 3,352 college students from 18 majors spanning 5 semesters and test it on another behavioral dataset from the subsequent grade of 3,245 students from 17 majors spanning 5 semesters. The evaluation results show the effectiveness of our proposed method on academic performance prediction. The results also demonstrate that each behavioral factor is effective for predicting academic performance.

The contributions of this paper are four-fold:
\begin{itemize}[leftmargin=*]
	\item Based on the big five personality traits, we quantify two personal behavioral factors (diligence and orderliness) and showed their strong correlation with academic performance.
	\item We measured students' behavioral similarity and incorporated it into our proposed prediction framework.
	\item We proposed a novel multi-task learning-to-rank model to predict students' academic performance.
	\item We conducted comprehensive experiments on a large-scale educational dataset. The results show the effectiveness of our proposed algorithm.
\end{itemize}

\section{Related Work}
\subsection{Analysis of Factors Influencing Academic Performance}
In the fields of education and psychology, much research has focused on identifying the predictors of college student's academic performance. Many existing studies concentrated on the association between students' personality traits, lifestyle behaviors (e.g., physical activity~\cite{gonzalez2004correlation,munoz2017influence}, sociability~\cite{robbins2004psychosocial}, sleep~\cite{dewald2010influence,zeek2015sleep}), intelligence level~\cite{laidra2007personality} and academic performance. Meta-analysis of the five-factor model of personality and academic performance indicated that academic performance is correlated with agreeableness, openness to experience, particularly conscientiousness~\cite{poropat2009meta,BARRICK:PEPS1}. \emph{However, almost all research is primarily based on students' self-reports and questionnaires, which may be susceptible to a range of limitations, such as small sample size, social desirability bias.}

\subsection{Academic Performance Prediction}
With the development of information technology, many efforts have been devoted to predicting performance based on students' digital records collected from online learning platform. For example, in an online learning environment, one line of research applied transfer learning~\cite{he2015identifying}, graphical model~\cite{fei2015temporal,wang2016nonlinear}, multi-view semi-supervised learning~\cite{li2016dropout} to predict student dropout by using their online behavior data. Another line of studies in online learning system utilized multiple instance learning was used in~\cite{zafra2011multiple}, tensor factorization~\cite{thai2011multi}, probabilistic latent semantic analysis~\cite{cetintas2013probabilistic}, fuzzy cognitive diagnosis framework~\cite{wucognitive} to predict student's performance. \emph{Different from previous studies on online learning platform, in this study, we mainly focus on offline behavior feature to predict academic performance.}

Based on the students' previous course records/grades and their demographical data, longitudinal data analysis is leveraged in~\cite{tamhane2014predicting} for predicting whether a student is at risk of getting poor assessment performance. Regression models~\cite{conijn2017predicting,arnold2012course}, support vector machine~\cite{huang2013predicting}, naive Bayes model~\cite{hien2007decision} are also utilized to predict students' academic performance by using historical course grades. \emph{However, temporal granularity of historical course performance is much coarser than campus behavior. In order to better provide timely personalized intervention and guidance according to campus behavior, the goal of our framework is to detect the effect of students' campus behavior on academic performance. These behavioral predictors are then used to predict academic performance.}

Recently, some studies started concentrating on predicting academic performance based on daily behavior data collected by the sensor. StudentLife~\cite{wang2014studentlife} and SmartGPA~\cite{rui2015smartgpa} studies found correlations between students' GPAs and automatic sensing behavior data obtained from smartphones. Another study~\cite{rettinger2011relationship} measured students' physical activity using a sensor armband in addition to self-reports and found that changes in physical activity were associated with GPA. However, the passive sensing behavior data they used is only collected from a small number of student volunteers, which may not be fully spontaneous. \emph{Different from previous work, our behavior data is left in the daily life on campus when using smart cards and our prediction is based on a large-scale behavior dataset which contains a large number of students.}
\section{Problem Statement}
In this section, we will introduce some notations and then formally define the problem in this work. In a university, let $\mathcal{M}=\{1, 2, ..., M\}$ and $\mathcal{S}=\{1, 2, ..., S\}$ denote the set of majors and semesters, respectively. The set of students in every major is defined as $Q=\{\mathcal{N}_1,\mathcal{N}_2,...,\mathcal{N}_M\}$. For every student $i$ in major $m$ at semester $s$, we define the feature vector and his academic performance as $\textbf{x}_i^{s,m}\in \mathbb{R}^p$ and $y_i^{s,m}$, respectively. Let $\textbf{X}^{s,m}=[\textbf{x}_{1}^{s,m}, \textbf{x}_{2}^{s,m}, ..., \textbf{x}_{|\mathcal{N}_m|}^{s,m}]^T\in \mathbb{R}^{|\mathcal{N}_m|\times p}$ and $\textbf{y}^{s,m}=[y_{1}^{s,m}, y_{2}^{s,m}, ..., y_{|\mathcal{N}_m|}^{s,m}]\in \mathbb{R}^{|\mathcal{N}_m|}$ denote the feature matrix and academic performance of all students in major $m$ at semester $s$. Note that, the academic performance in this paper is represented by the students' rank. The details of features will be described in the following section. Then we formally define our academic performance prediction problem as follows:

\textbf{Academic Performance Prediction Problem}: At semester $s$, given the feature matrices $\textbf{X}_i^{s,1}$, $\textbf{X}_i^{s,2}$, ..., $\textbf{X}_i^{s,M}$ of every major, we are supposed to predict the corresponded academic performance $\textbf{y}_i^{s,1}$, $\textbf{y}_i^{s,2}$, ..., $\textbf{y}_i^{s,M}$ in this semester.  
\section{Behavioral Analysis}
In this section, we will first introduce the dataset for this problem, and then analyze the relation between behavioral features (i.e., diligence, orderliness, sleep pattern) and academic performance. Finally, we measure and analyze student similarity based on social influence theory. Such analysis motivates us to build a meaningful academic performance prediction model.
\subsection{Dataset}
We collect data from one university during 2011/09/01 to 2015/06/30. This dataset consists of two types of data, which are described as follows:
\subsubsection{Behavioral Data}
In most universities, every student owns a campus smart card as the recognition tool for his identification. These smart cards could be used for the unique payment (\emph{access}) medium for many consumptions (\emph{resources}), and thus record a large volume of behavioral data, including almost all potential activities on campus, such as fetching boiled water in the teaching building, entering the library and paying for meals in the cafeteria. The collection of smart cards data is unobtrusive, which is different from previous studies whose data is mainly collected from self-reports and questionnaires, and suffer from a small sample size and social desirability~\cite{vedel2014big,poropat2009meta,BARRICK:PEPS1,crede2010class,wald2014associations}.
\subsubsection{Academic Performance Data}
Students' academic performance is also be recorded, which contains the grade, credit of each course for each student. For protecting privacy, in this paper, we calculate Grade Point Average (GPA) for each student and then get the rank of GPA. Finally, in each semester, we normalize the ranking list to $[0,1]$ within every major and get the normalized rank list. The normalized rank is regarded as the academic performance of each student, whose value close to zero means better performance.

\subsection{Personality Analysis}
Big Five personality traits (i.e., Openness, Conscientiousness, Extraversion, Agreeableness, Neuroticism) is a traditional psychology model based on common language descriptors of personality~\cite{gosling2003very}. Previous work has shown the effect of the five-factor model on job/academic performance (conscientiousness in particular)~\cite{vedel2014big,poropat2009meta,BARRICK:PEPS1}. Based on the usage data of smart cards and motivated by this effect, we first extract two types of important predictors: diligence and orderliness, and then analyze their correlation with academic performance. 

\subsubsection{Diligence}
Diligence is strongly related to a narrow trait of conscientiousness: achievement, which reflects the tendency to strive for competence and success in ones' work/study~\cite{zulauf1999use}. From the usage data of smart cards, diligence is mainly represented by the amount of time that students spend on their study. Although we couldn't have an explicit measure, we count the occurrence frequencies at learning areas as a proxy. The most evident study areas include the library and teaching buildings. In consideration of safety, libraries in modern universities establish entrance guard systems so that students are required to swipe their smart cards before entering libraries. Besides, the library provides all kinds of books for students to borrow but requiring their smart cards for authorization. In the teaching building, although there are no such facilities, students may fetch boiled water for drinking when they attend classes or review lessons, particularly in winter. For the sake of water conservation, students also need to make payment, in spite of only involving a few cents. These payment records can be regarded as the proxy of studying in the teaching building.
\begin{figure}[h]
	\centering
	\subfigure[library v.s. rank]{
		\centering
		\label{fig:diligence:a}
		\includegraphics[width=0.35\textwidth]{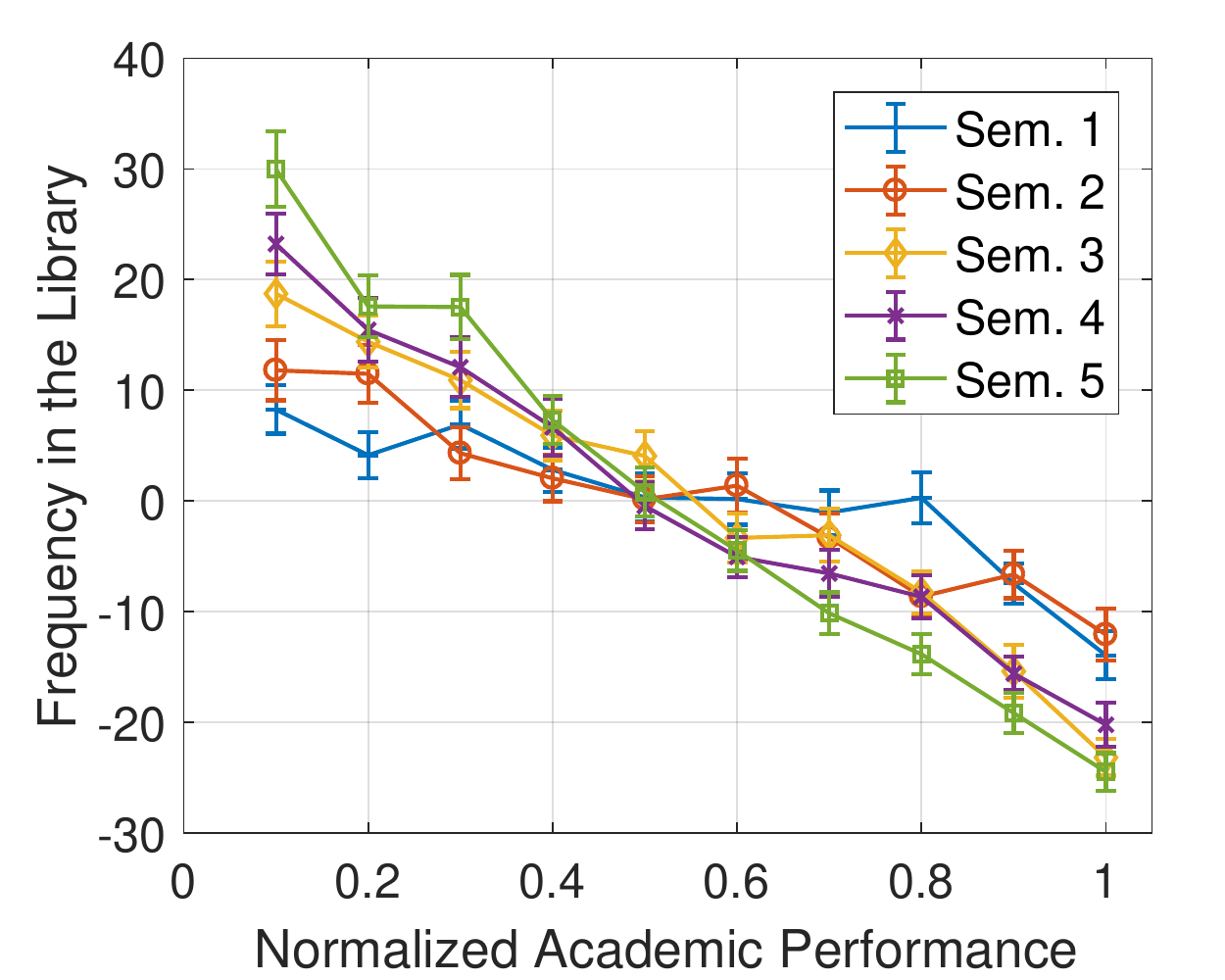}}
	\subfigure[teaching building v.s. rank]{
		\centering
		\label{fig:diligence:b}
		\includegraphics[width=0.35\textwidth]{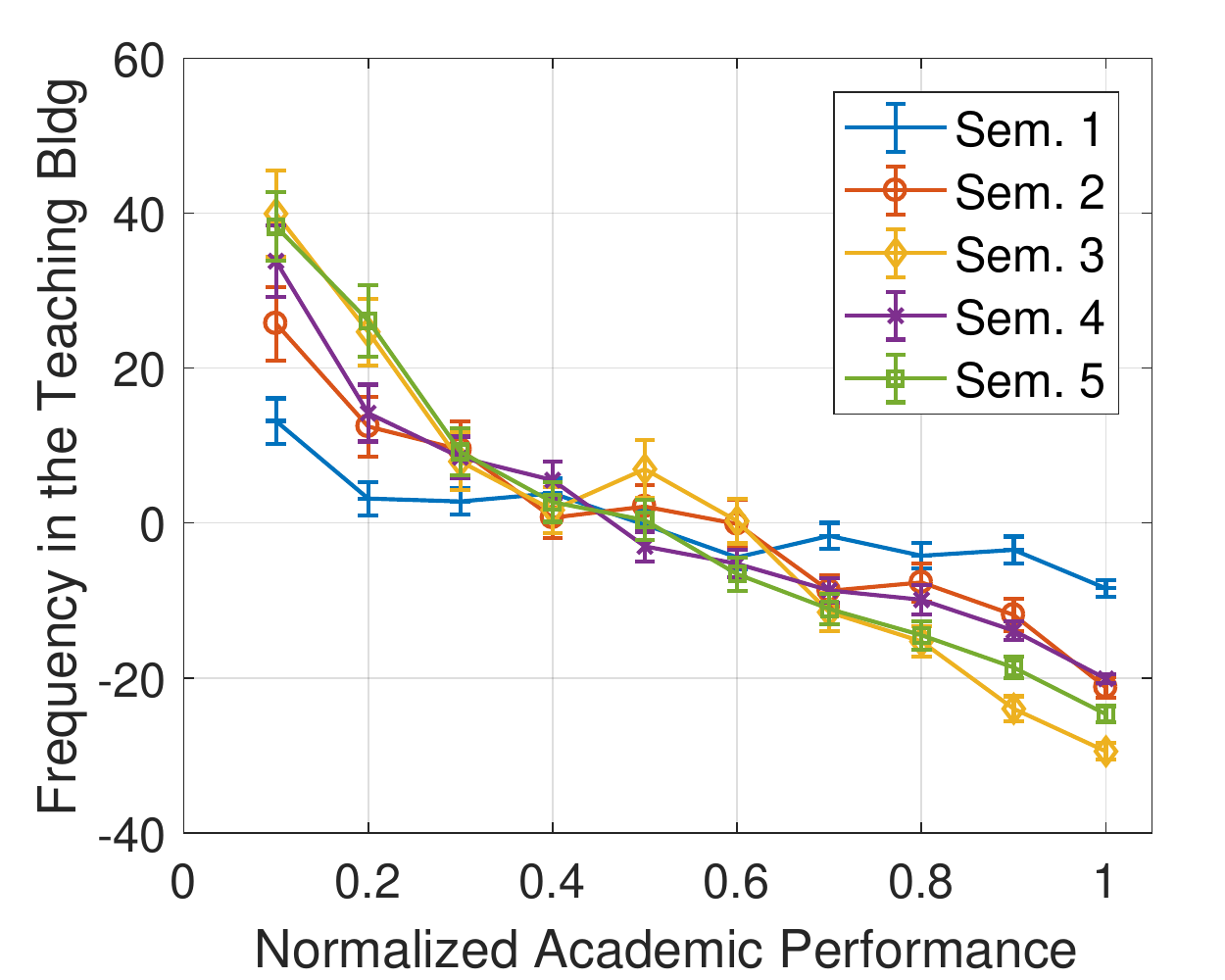}}
	\caption{The illustration of correlation between diligence and academic performance.} 
	\label{fig:diligence_all}
\end{figure}

Therefore, for diligence, we calculate the occurrence frequency at the library and teaching building. The correlation of these diligence factors with academic performance is shown in Figure~\ref{fig:diligence:a} and Figure~\ref{fig:diligence:b}, where we remove the influence of mean value of diligence for comparison across semesters. The higher value of occurrence frequency at the library and teaching building means the higher degree of diligence. We observe that diligence is negatively correlated with the performance rank, where the averaged Spearman correlation over semesters can achieve as much as $-0.308$. Such an observation indicates student's hard study can be paid back, harvesting good academic performance. Besides, we also discover that the correlation varies from semester to semester. Particularly, the Spearman correlation in the first semester is significantly lower than the other semesters, whose value are $-0.176$ and $-0.157$ in Figure~\ref{fig:diligence:a} and Figure~\ref{fig:diligence:b}, while the correlation of other semesters are higher than $-0.249$. One potential reason is that academic performance in the first semester still highly depends on the knowledge obtained in high school and there are no large behavioral differences between students. The correlation of more diligence factors with academic performance has been shown in Table~\ref{tab:feature}. It is worth noting that we have put the frequencies of going to printing room into diligence characteristics because they need many teaching materials for reviewing before the exam. 

\subsubsection{Orderliness}
Orderliness is another important narrow trait of conscientiousness, which reflects the tendency to keep things organized and tidy~\cite{dudley2006meta} and has been reported another important factor that can affect academic performance~\cite{higgins2007prefrontal,poropat2009meta,BARRICK:PEPS1,cao2018orderliness}. Orderliness can be measured as the regularity of activities, such as having breakfast, taking a shower, etc. In general, the regularity of each activity can be quantified as the Shannon entropy of its temporal distribution, which is defined as:
\begin{equation}
H=-\sum_{i}p(i)\log p(i).
\end{equation}
Where $p(i)$ is the probability of carrying out the activity during the $i^{\text{th}}$ period of a day and estimated by kernel density estimation. In our case, each period is defined as one hour. The entropy $H$ represents the temporal uncertainty of the activity. 
We use entropy for measuring regularity of taking a shower and shopping. In our dataset, all students have the records of taking shower and almost 95\% students go to the supermarket in one semester, so these two activities are high-frequent and can objectively reflect students' lifestyle. The larger the entropy, the higher the temporal uncertainty of these two activities, i.e., the lower their orderliness. 

Breakfast is another important activity that reflects the regularity of students' life. However, the time of having breakfast is easily affected by the time of the first class each day, which differs from day to day for students of the same major. 
Thus, for measuring the regularity of having breakfast, we only compute the frequency that each student swipes his card in the cafeteria from 6 am to 9 am. Almost 98\% students have records in this time period. The larger the frequency, the higher the regularity of having breakfast. 

\begin{figure}[h]
	\centering
	\vspace{-1em}
	\subfigure[breakfast order v.s. rank]{
		\centering
		\label{fig:order:a}
		\includegraphics[width=0.35\textwidth]{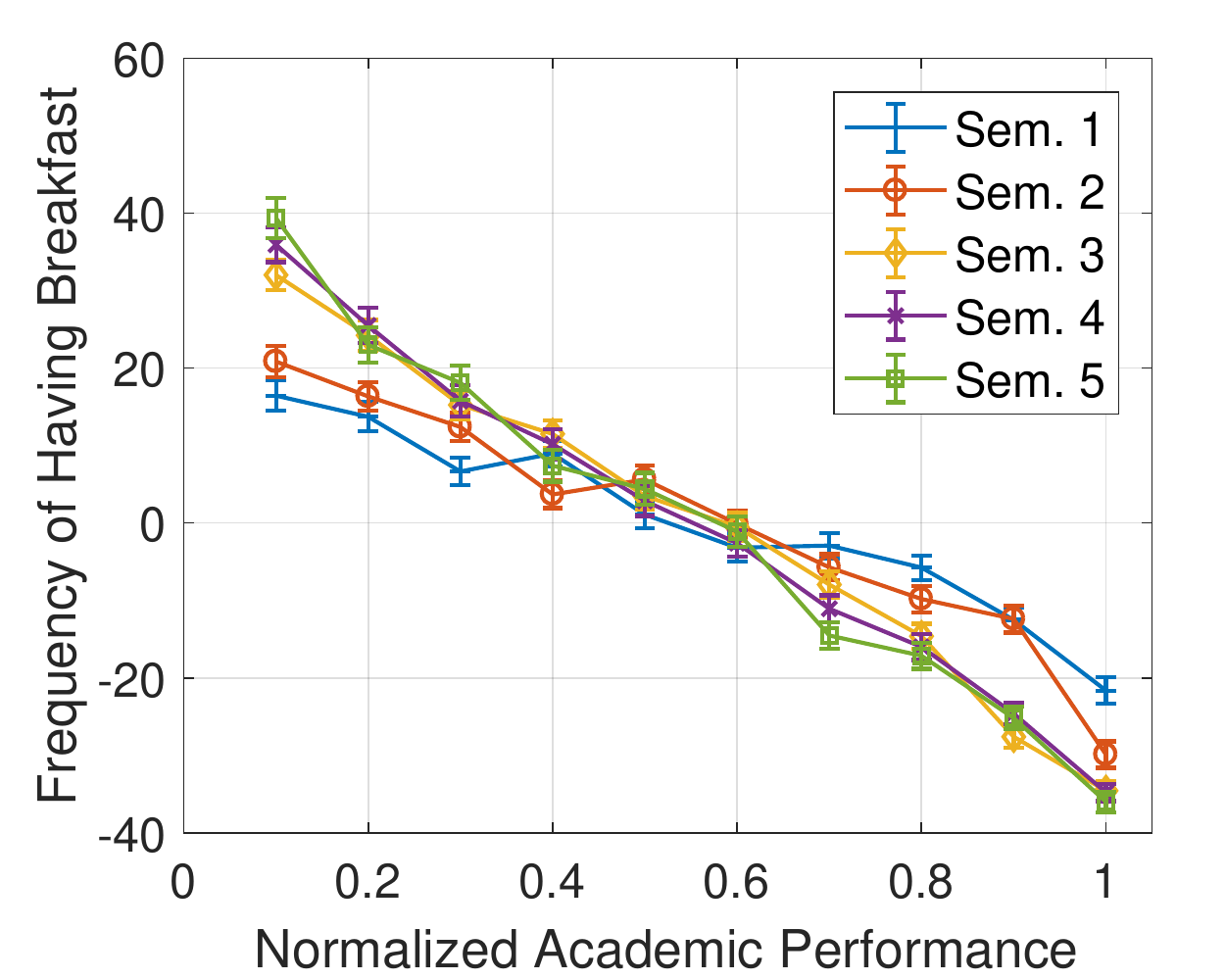}}
	\subfigure[shower order v.s. rank]{
		\centering
		\label{fig:order:b}
		\includegraphics[width=0.35\textwidth]{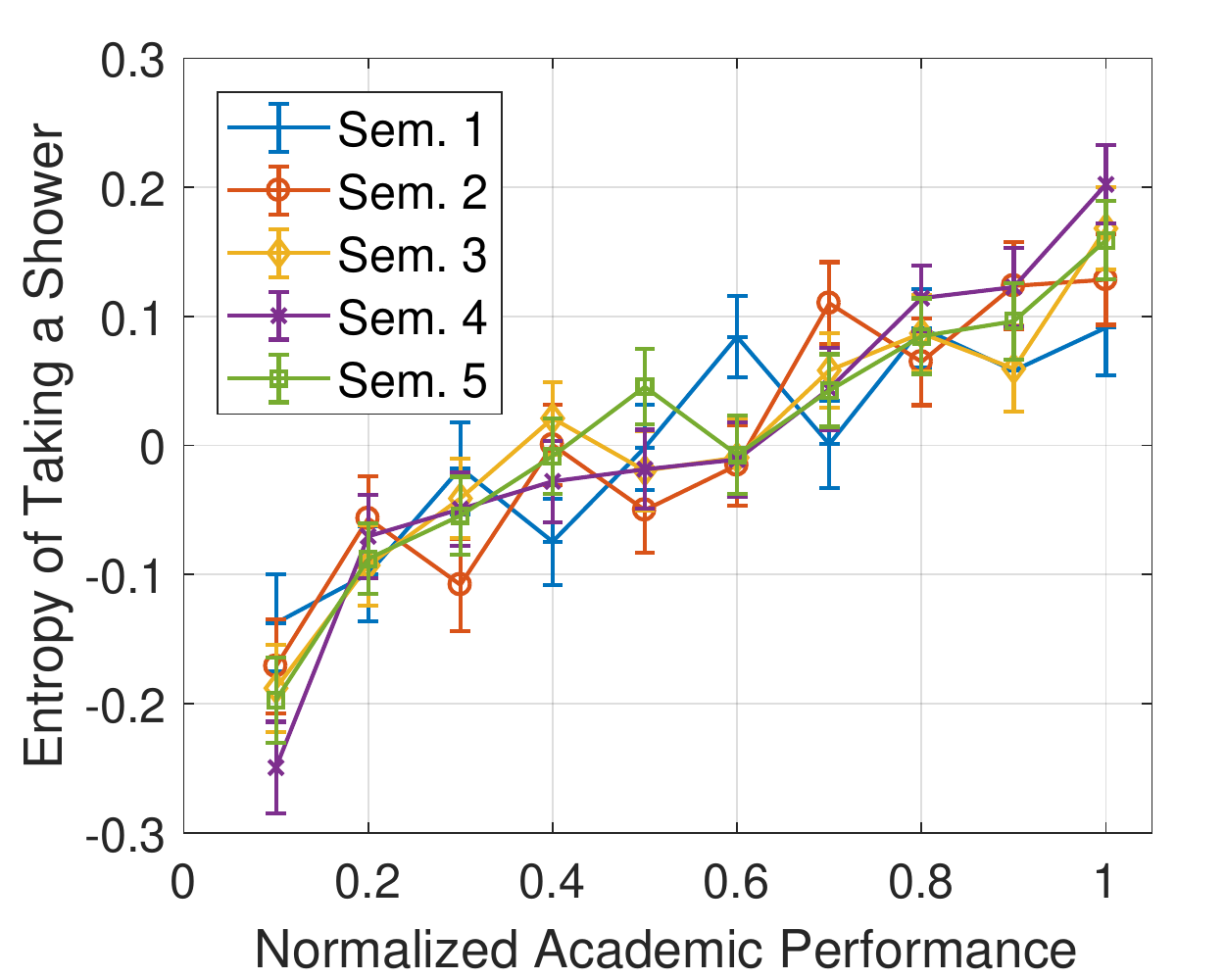}}
	\caption{The illustration of correlation between orderliness and academic performance. }
	\label{fig:order_all}
	\vspace{-1em}
\end{figure}

The correlations of the regularity of breakfast and shower with the performance rank are shown in Figure~\ref{fig:order:a} and Figure~\ref{fig:order:b}. The effect of the mean value of orderliness is also removed. These two figures indicate that regularity is positively correlated to the academic performance.
Besides, similar to the result of diligence, the correlation of orderliness varies among different semesters and thus can be explained by the same underlying reasons. The correlation of shopping regularity with performance rank is also computed and shown in Table~\ref{tab:feature}.
\subsection{Sleep Pattern Analysis}
According to the previous studies about the correlation between students' sleep patterns with their academic performance~\cite{trockel2000health,dewald2010influence}, students with good sleep habits tend to achieve better performance. In particular, the wake-up and bedtime time are of great importance for academic performance~\cite{trockel2000health,taylor2013role}, students with later wake-up time and bedtime tend to perform worse.

\begin{figure}[h]
	\vspace{-1em}
	\centering
	\subfigure[wake-up time]{
		\centering
		\label{fig:wakeup}.
		\includegraphics[width=0.35\textwidth]{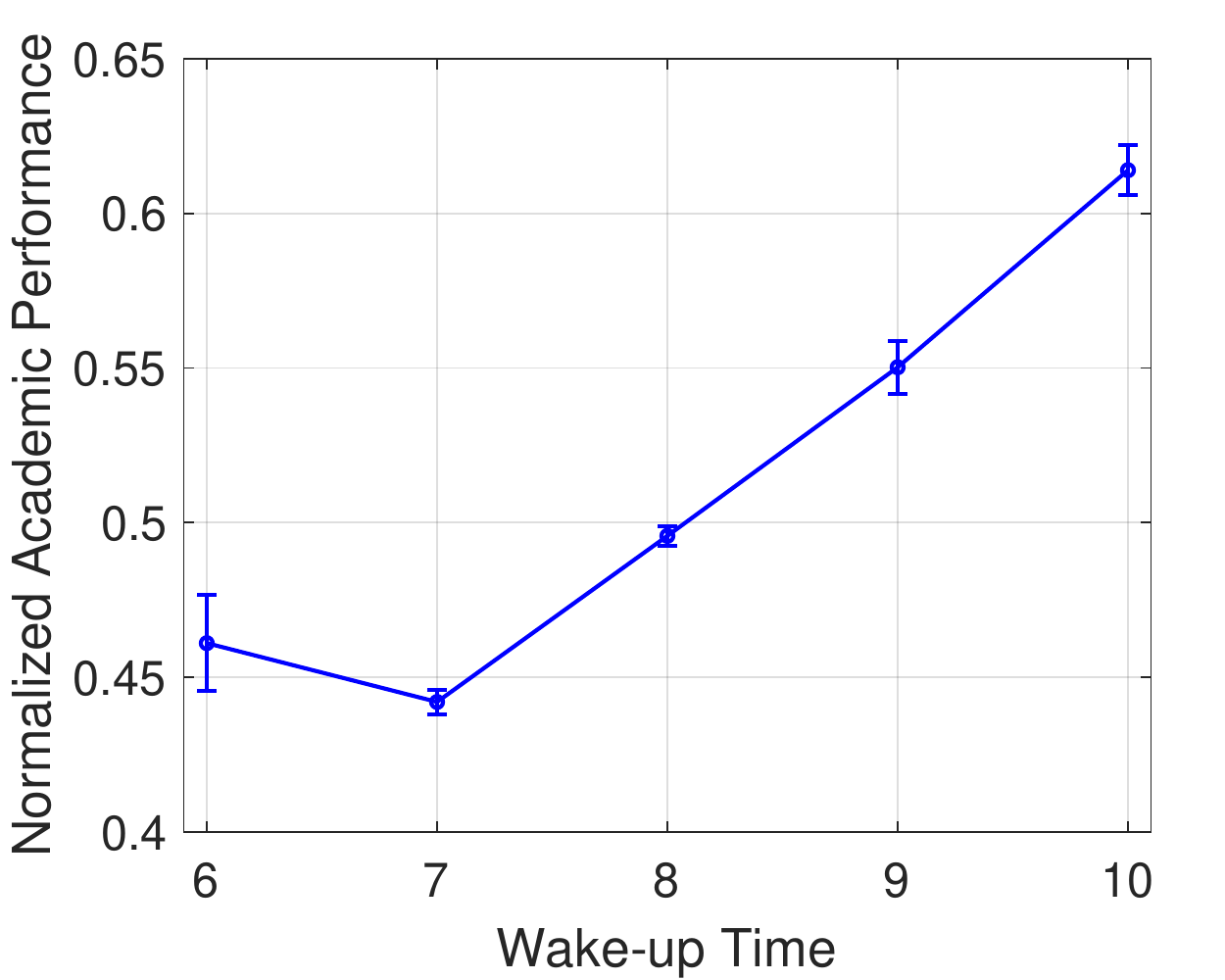}}
	\subfigure[bed time]{
		\centering
		\label{fig:bed}
		\includegraphics[width=0.35\textwidth]{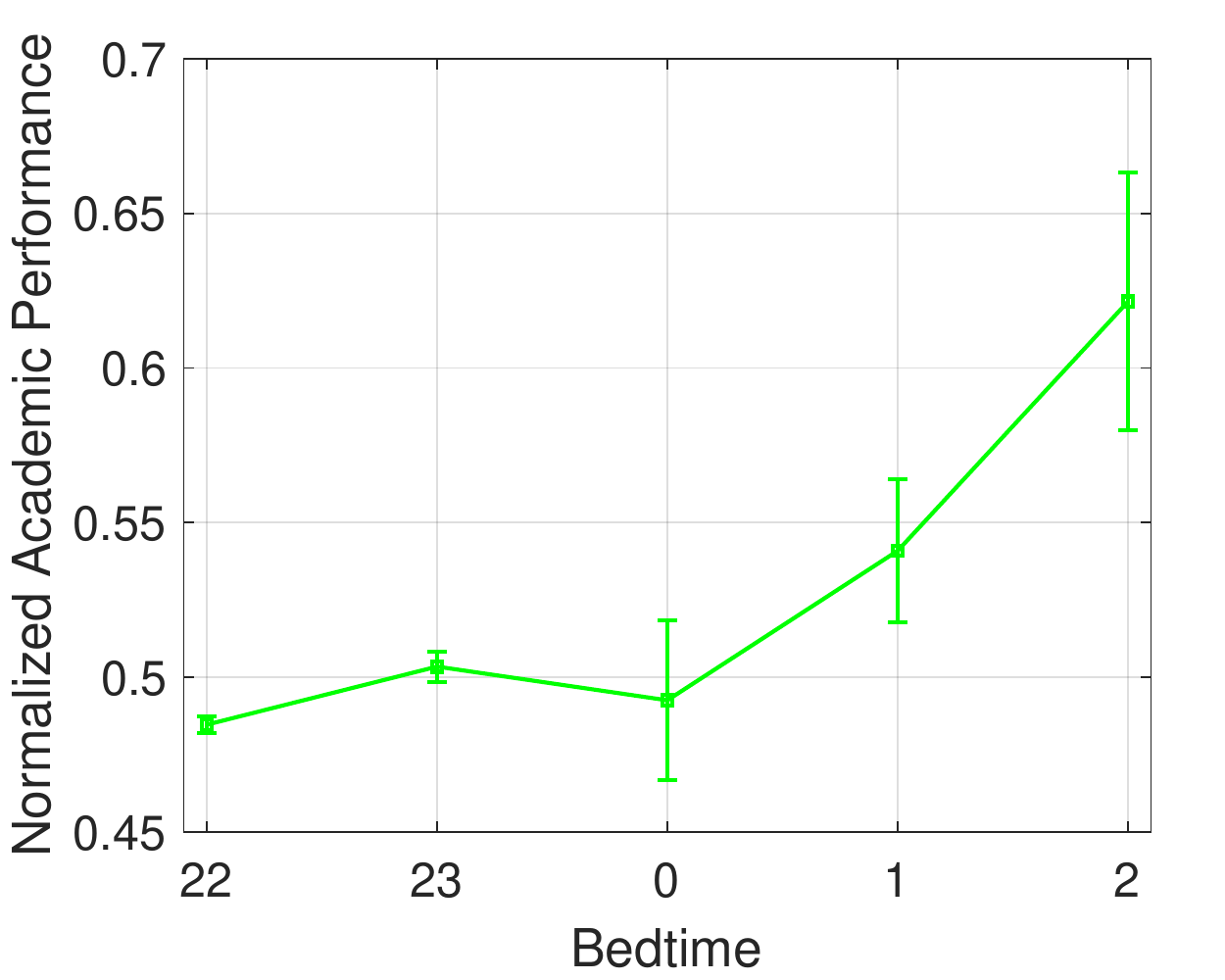}}
	\caption{The illustration of correlation between wake-up and bed time. In (b), bed time with value 0, 1, 2 means students go to bed in the next day.}
	\vspace{-1em}
\end{figure}

In our dataset, we do not have the actual information of bedtime and wake-up time. However, students need to use their smartcard to fetch the hot water, to enter-exit their dormitories and to take a shower in the morning or evening. Thus, the last and first smart card records in each day are regarded as surrogates of students' wake-up time and bedtime respectively. Considering the different starting time of classes in the morning and end time in the evening, too specific wake-up and bedtime may not represent the actual sleep pattern. So we calculate the frequency of first and last hours of smartcard records and choose the timestamp of highest frequency, which are used to represent each student's sleep pattern. We only have several concrete values (e.g. wake-up time concentrates in 6, 7, 8, 9, 10, 6 means students wake up at 6:00 am - 7:00 am). The correlation of each student's academic performance under different wake-up and bedtime are shown in Figure~\ref{fig:wakeup} and Figure~\ref{fig:bed}, respectively.

In Figure~\ref{fig:wakeup}, students who wake up late may be associated with poorer performance. Since the earliest classes usually start at 8:30 am, students wake up at 6:00 am and 7:00 am are both considered as the early wake-up, and they achieve the best performance. In Figure~\ref{fig:bed}, we can see students who go to bed too late (after 1:00 am) perform worse, which may be explained by the addiction of online games. In spite of showing interesting correlation with academic performance, sleep patterns do not take significant effect, after being integrated with diligence and orderliness. One potential reason may lie in bias estimation of sleep/wake-up time. The features of sleep patterns are also shown in Table~\ref{tab:answer}. Note that, sleep pattern features are one-hot features, so we cannot calculate the Spearman correlation. Instead, we discretize the academic performance to 5 groups. Then, we calculate the Cram\'{e}r's V value~\cite{cramer2016mathematical} between discretized academic performance and sleep pattern as the correlation coefficient and show the results in Table~\ref{tab:feature}.
\begin{table}[h]
	\centering
	\caption{Student Behavioral Features. The correlation coefficient of sleep pattern is Cram\'{e}r's V value, while others are Spearman value. All p-value of correlation is greatly smaller than 0.001. ``Before exams'' means 20 days before final exams.}
	\begin{tabular}{l|l|l}
		\hline
		& Features & Correlation Coefficient\\\hline
		\multirow{9}{*}{Diligence}
		& freq. in the lib & -0.3063\\
		& freq. in the lib at weekends & -0.3262\\
		& freq. in the lib before exams & -0.2984\\
		& freq. of borrowing books & -0.3031\\
		& freq. of fetching water & -0.3258\\
		& freq. of fetching water at weekends & -0.2997\\
		& freq. of fetching water before exams & -0.3308\\
		& freq. of printing & -0.1324\\
		& freq. of printing before exams & -0.0943\\\hline
		\multirow{3}{*}{Orderliness}
		& breakfast & -0.3621\\
		& shower & 0.1697\\
		& shopping & 0.0809\\\hline
		\multirow{2}{*}{Sleep$^1$}
		& bed time & 0.0589\\
		& wake-up time & 0.0765\\\hline
	\end{tabular}
	\vspace{0.3em}\\
	$^1$\footnotesize{Sleep means sleep patterns.}
	\label{tab:feature}
\end{table}
\subsection{Student Similarity}
As suggested in~\cite{yao2017predicting}, two students may have similar academic performance if they behave in a similar way. Student similarity is another important factor to analyze students' academic performance, which is measured by their co-occurrence in this paper. Each co-occurrence is assumed as two students generating records at the same location within a short time interval, which is empirically set as one minute. However, some co-occurrence may occur by random, which motivates us to set a threshold to remove the effect of these random cases.

\begin{figure}[htb]
	\vspace{-1em}
	\centering
	\subfigure[cafeteria]{
    	\centering
    	\label{fig:sto_mess}
    	\includegraphics[width=0.3\textwidth]{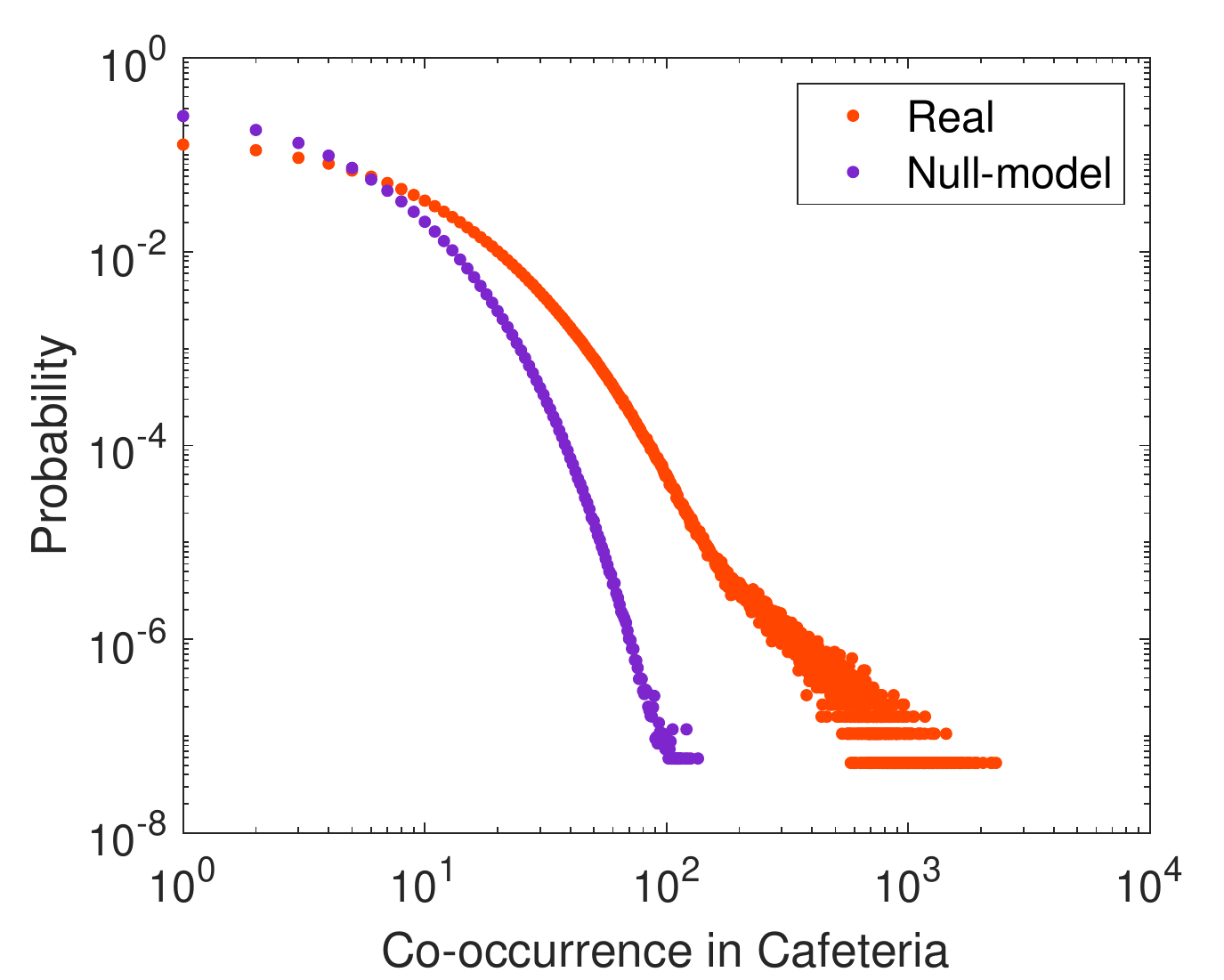}}
	\subfigure[supermarket]{
		\centering
		\label{fig:sto_super}
		\includegraphics[width=0.3\textwidth]{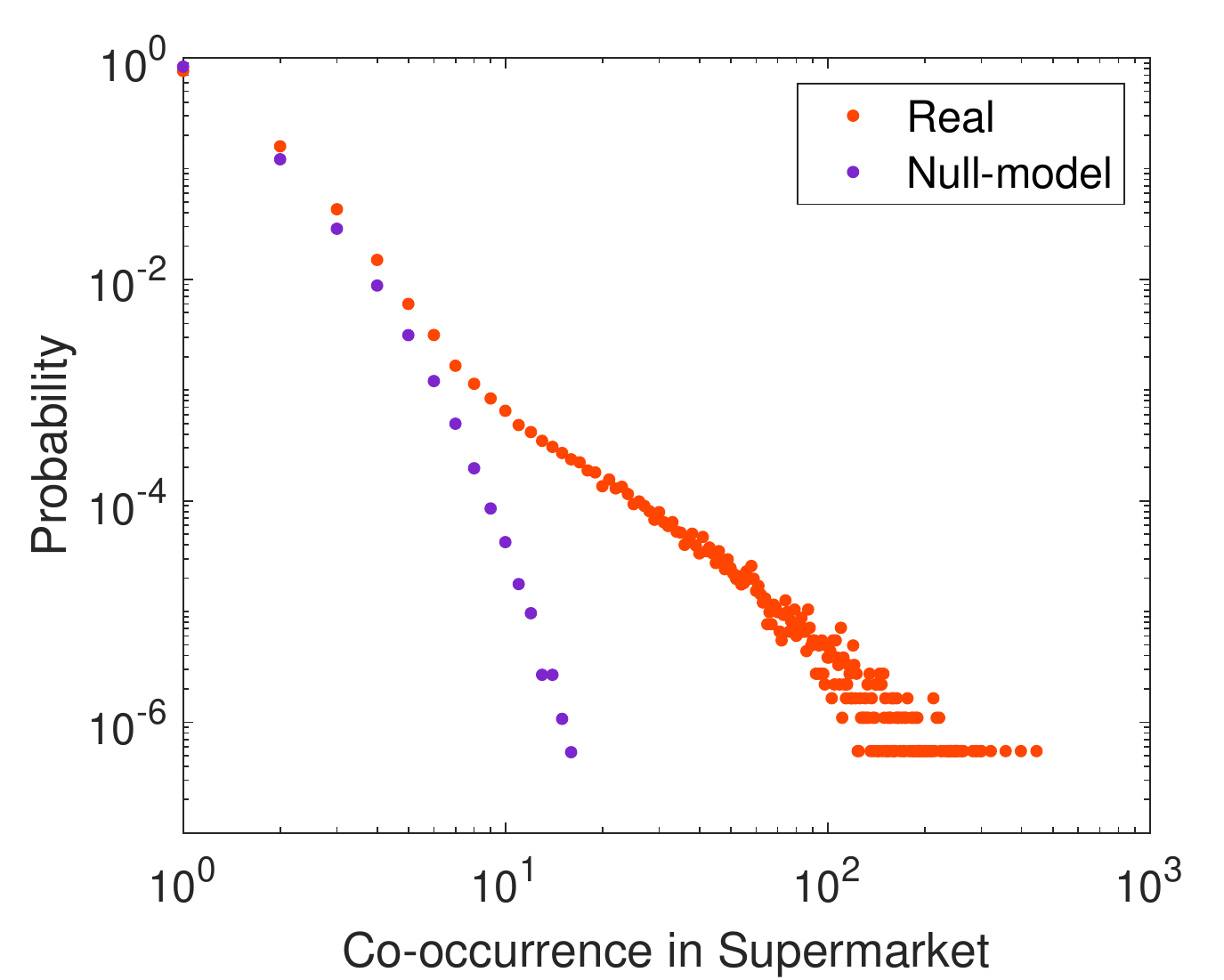}}
	\subfigure[library]{
		\centering
		\label{fig:sto_lib}
		\includegraphics[width=0.3\textwidth]{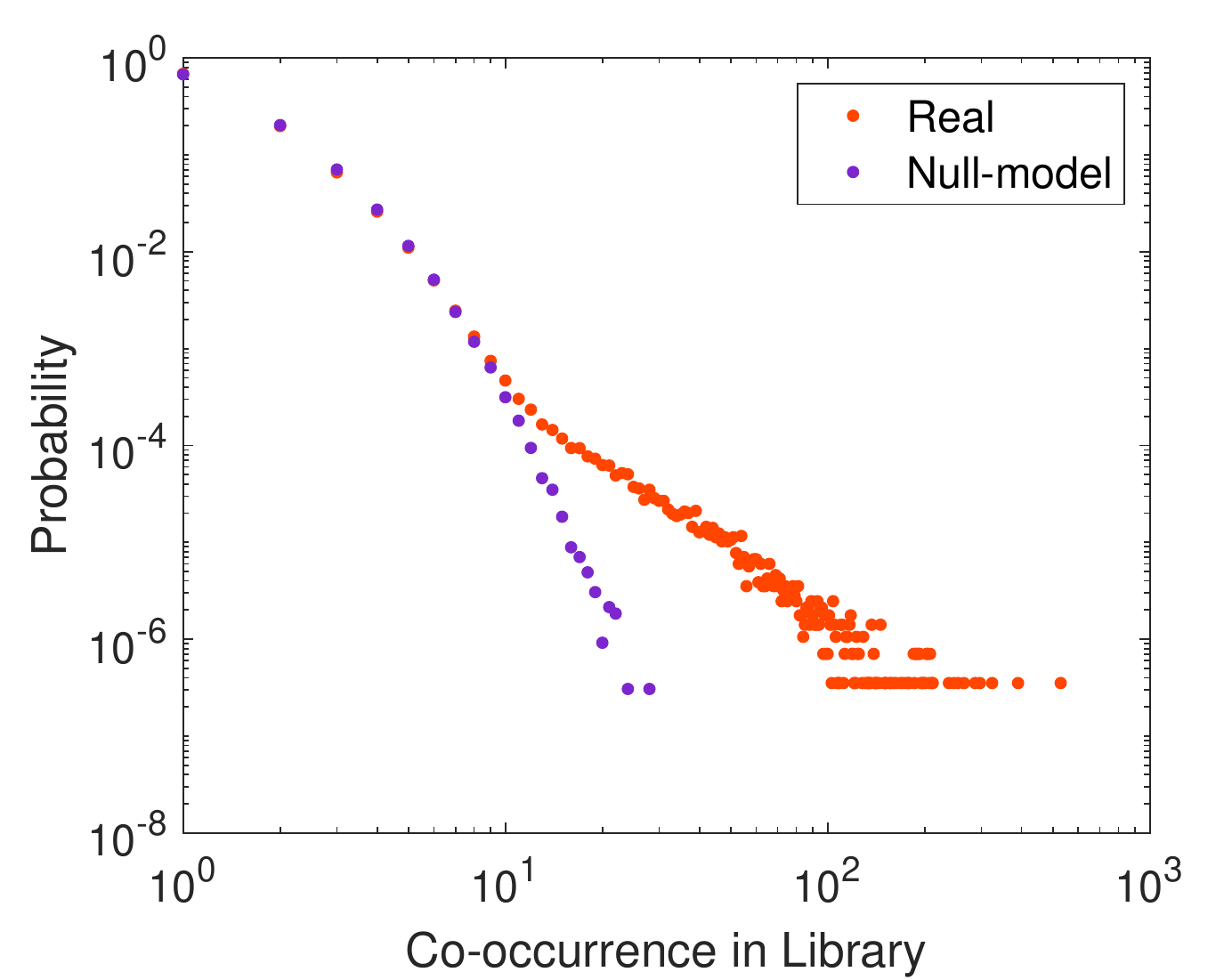}}
	\caption{(a) (b) (c) represent the comparison of co-occurrence distribution of null model and real case at cafeteria, supermarket and the library respectively.}
	\vspace{-1em}
\end{figure}

Follow the settings in~\cite{yao2017predicting}, we first construct a null model by randomly shuffling the timestamps of behavioral records. For example, if one student's record of location-time pair is \emph{\{(21:50:55, teaching building); (22:03:06, library); (23:00:05, dormitory)\}}, we shuffle these three timestamps. Then we compute the mean and standard deviation of co-occurrence frequency by repeating the construction process 20 times. The comparison of the co-occurrence frequency between the null model and the real case at cafeteria, supermarket and library are shown Figure~\ref{fig:sto_mess}, Figure~\ref{fig:sto_super} and Figure~\ref{fig:sto_lib}, respectively. We determine the threshold by keeping the co-occurrence frequency of the real case above the mean co-occurrence frequency plus two times of standard deviation of the random case, and in these three locations is $18$, $112$ and $23$, respectively. 

Based on the derived threshold, we drop those co-occurrence frequencies below the threshold at each location. Furthermore, in this study, the co-occurrence frequencies are computed from $L$ different locations, so we need to combine these frequencies together. However, since we have neither evidence on the importance of different locations nor real similarity information used as training data for location importance learning, we simply define the similarity between student $i$ and student $j$ as:
\begin{equation} 
\label{eq:simi}
\tau_{ij}=\sum_{l=1}^{L}\tau^{l}_{i,j}/\max_{j}\tau^{l}_{i,j},
\end{equation}
where $\tau^{l}_{i,j}$ is the co-occurrence frequency at the location $l$. 

\begin{figure}[h]
	\centering
	\includegraphics[width=0.55\textwidth]{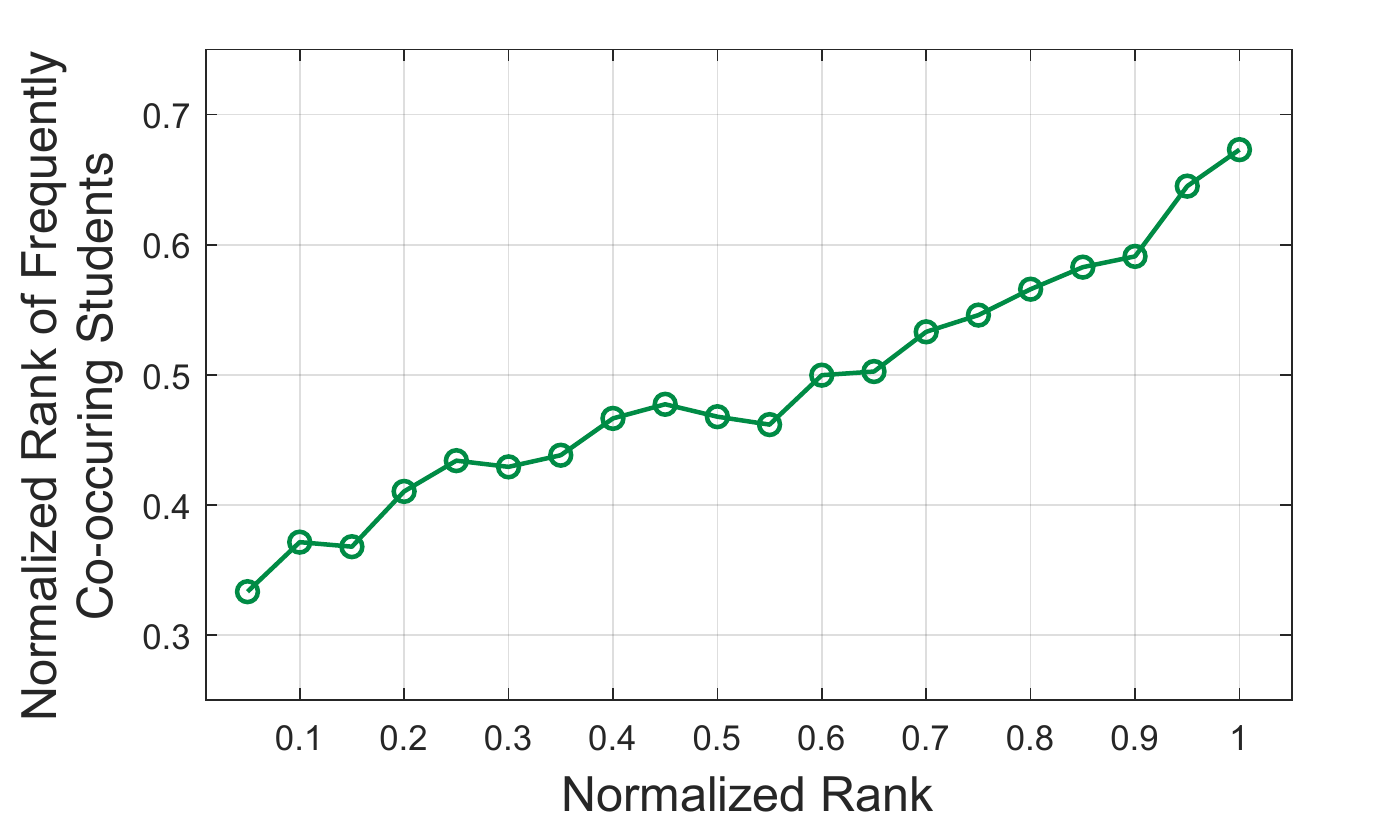}
	\caption{Correlation between students' performance and the average of their similar students' performance}
	\label{fig:friend}
\end{figure}

As we mentioned before, if two students behave similarly, they may have similar academic performance. In other words, each student's academic performance should be close to those who frequently co-occurring with him. For each student $i$, we can get one student group $F_i$ contains students who frequently co-occur with $i$.
Based on our dataset, we construct a campus social network and the result of this study is shown in Figure~\ref{fig:friend}. 
The x-axis represents the rank of each student $i$ and the y-axis represents the average rank of students in the corresponding student group $F_i$. The Spearman correlation between students' academic performance and the averaged academic performance of students in the corresponding group is $0.434$. 

Furthermore, we test the effect of student similarity with regard to academic performance by comparing the similarity between similar student group with nonsimilar student group, Similar as~\cite{gao2012exploring,yao2017predicting}, we first define the similarity of academic performance between a student $i$ and a student group $R$ as:
\begin{equation}
Q_R(i)=\frac{\sum_{j\in R}sim(i,j)}{|R|}, 
\end{equation}
where $sim(i,j)=|y_i^{s,m}-y_j^{s,m}|$ represents the similarity of academic performance between student $i$ and $j$. The average similarity between student $i$ and his similar students and the average similarity between student $i$ and bootstrap-sampled 20 students from his non-similar students' list are calculated and defined as $\mathbf{Q}_F(i)$ and $\mathbf{Q}_{NF}(i)$. We then conduct the two sample t-test on the vector $\mathbf{Q}_F$ and $\mathbf{Q}_{NF}$, where the null hypothesis is $H_0: \mathbf{Q}_{NF}\neq \mathbf{Q}_F$ and the alternative hypothesis is $H_a: \mathbf{Q}_{NF}>\mathbf{Q}_F$. The number of students (i.e., the size of vector $\mathbf{Q}_{F}$ and $\mathbf{Q}_{NF}$) is 3,245. The degrees of freedom is 6,488 and t-statistic is -28.93. The null hypothesis is rejected at significant level $\alpha= 0.001$ with p-value$<0.0001$, which indicates similar students have similar academic performance.
\section{Proposed MTLTR-APP Framework}
In this section, we provide details for our proposed \textbf{M}ulti-\textbf{T}ask \textbf{L}earning-\textbf{T}o-\textbf{R}ank \textbf{A}cademic \textbf{P}erformance \textbf{P}rediction (MTLTR-APP) framework. First, we introduce the basic pair-wise learning-to-rank model. Then, we introduce how to model intra-major, intra-semester and student similarity into an optimization framework. Finally, we discuss how to optimize the framework and how to utilize the model to predict academic performance.
\subsection{Basic Pair-wise Learning-to-Rank Model}
According to behavioral analysis, we extract the three behavioral factors $\textbf{x}^{s,m}_i\in \mathbb{R}^p$ listed in Table~\ref{tab:feature} in semester $s$ for each student $i$ in major $m$. Based on these factors, we present the proposed framework, named as RLTR-SEQ for predicting students' academic performance. The algorithm of academic performance prediction is based on the pair-wise learning-to-rank model, which maps a pair of features into their relative performance. The feature scoring function is defined as $f(\textbf{x}_i)=\mathbf{w}^{s,m}_i\textbf{x}^{s,m}_i$. Then, the pair-wise mapping function is defined as:
\begin{equation}
P(i\triangleright j) = \sigma(f(\textbf{x}_i)-f(\textbf{x}_j)),
\end{equation}
where $i\triangleright j$ indicates the student $i$ should be predicted to outperform the student $j$. $\sigma(x)=1/(1+e^{-x})$ is a Sigmoid function. 

Additionally, the difference of the correlations between each behavior factor and the academic performance among different semesters indicates the factors' district predictive strength (e.g. the correlations between diligence and academic performance are shown in Fig.~\ref{fig:diligence_all}). Thus, in this framework, we regard the prediction at a specific semester of one major as a task. As we described before, there are $S$ semesters and $M$ majors. Furthermore, we exploit the cross-entropy to measure the inconsistency of the hypothesis and the real relative performance. Since the performance rank is obtained based on GPA, it is a rare case that two students have the same rank (i.e., rank tie). Without taking the rank tie into account, the basic loss function of learning-to-rank $\mathcal{L}_{RN}$ is then represented as:
\begin{equation}
\label{eq:ranknet}
\mathcal{L}_{RN}=-\sum_{s=1}^S\sum_{m=1}^M\sum_{y_i^{s,m}>y_j^{s,m}}\log\sigma(\textbf{w}^{s,m} \textbf{x}_i^{s,m}-\textbf{w}^{s,m} \textbf{x}_j^{s,m}),
\end{equation}
where $y_i^{s,m}$ and $y_j^{s,m}$ means the real performance of student $i$ and student $j$, and $y_i^{s,m}>y_j^{s,m}$ indicates student $i$ outperforms than student $j$ in a specific task. 
\subsection{Modeling Inter-semester Temporal Correlation}
According to the Figure~\ref{fig:diligence_all} and Figure~\ref{fig:order_all}, the correlation coefficients between the predictors and academic performance are gradually changed as time goes by. The findings motivate us to model the inter-semester temporal correlation. Therefore, we impose a sequentially smoothed regularization for the gradual change of weight, which is formally defined as:
\begin{equation}
\label{eq:temporal}
\Omega_{SEQ} = \frac{1}{2}\sum_{s=1}^S\sum_{p>s}^K \mathcal{A}(\Delta t)\Vert \textbf{W}^{s} - \textbf{W}^{p} \Vert_F^2,
\end{equation}
where $\Delta t=p-s$, $\textbf{W}^s=[\textbf{w}^{s,1}; \textbf{w}^{s,2}; ...; \textbf{w}^{s,M}]\in \mathbb{R}^{M\times p}$ is the weight matrices of semester $s$. $\mathcal{A}(\Delta t)$ is a decay function of $\Delta t$. $\Vert \cdot \Vert_F$ is the Frobenius norm of matrix. In this work, we discuss a special case of $\mathcal{A}(\Delta t)$. $\mathcal{A}(\Delta t)=1$ when $\Delta t=1$, otherwise $\mathcal{A}(\Delta t)=0$. Then Eq.~\eqref{eq:temporal} can be reformulated as:
\begin{equation}
\Omega_{SEQ} = \frac{1}{2}\sum_{s=1}^{S-1}\Vert \textbf{W}^{s} - \textbf{W}^{s+1} \Vert_F^2
\end{equation}
More specially, the weight of a major at the semester $s$ is actually learned by adding a zero-mean Gaussian distributed offset on the weight of the preceding semester. In other words, the regularization term can control the weight of parameter to avoid some sudden changes result from potential outliers among consecutive semesters, which can make the model more robust.

\subsection{Modeling Inter-major Correlation}
In our dataset, the number of students varies from major to major, the amount of training data in some tasks is limited, especially in a pairwise case whose sample number is $N^2$. In our dataset, the biggest major has about 600 students and the smallest major only has 50 students, which is corresponded to 360,000 and 2,500 samples. As a consequence, it is necessary to leverage the relationship between major-specific tasks to alleviate the sparsity problem of some tasks. 

In our problem, the relationship between majors may be explained by the similarity of courses and teaching styles from similar majors, such as electronic engineering and computer science. Students from these two majors have several common courses and are required excellent programming skills. For each semester, we assume that students from different majors within the same category (e.g., social science, engineering) have similar behavior. Thus, the major-related model can be factorized as the product of two low-rank matrices, i.e., $\textbf{W}^s=\textbf{U}^s\textbf{V}^s$, where $\textbf{U}^s\in \mathbb{R}^{M\times k}$ and $\textbf{V}^s \in\mathbb{R}^{k\times p}$. $\textbf{U}^s$ represents the weighted combination of categories to form the major. $\textbf{V}^s$ denotes the feature representation of these categories. The sparsity constraint of $\textbf{U}^s$ ensures that each major-related task depends on a small set of categories. In addition, we add a non-negative constraint on the matrix $\textbf{U}^s$ to ensure the interpretability. Formally, the regularization of inter-major correlation can be defined as:
\begin{equation}
	\begin{aligned}
		&\Omega_{MS}=\sum_{s=1}^{S}\lambda_1\Vert \textbf{U}^s\Vert_1+\lambda_2\Vert \textbf{V}^s\Vert_2^2\\
		&s.t., \textbf{W}^s=\textbf{U}^s\textbf{V}^s,\;\textbf{U}^s\succeq \textbf{0}
	\end{aligned}
\end{equation}
where $\lambda_1$ and $\lambda_2$ are regularization parameters, $\textbf{U}^s\succeq \textbf{0}$ means $\textbf{U}^s$ should be non-negative. Note that, for each major $m$, $\textbf{w}^{s,m}=\textbf{u}^{s,m}\textbf{V}^{s}$.

\subsection{Modeling Behavior Similarity}
In our dataset, some students do not have enough behavioral data. For example, some students have takeaway food more frequently than having food in the cafeteria, others may study more in the dormitories rather than the library or teaching buildings. These data can not be collected by the smart card, which may result in the unreliability of predictors. To address this challenge, based on the similarity of academic performance between frequently co-occurring students, students' predicted performance should not have large discrepancy from their similar students. we propose a regularization term defined as follow: 
\begin{equation}
	\label{eq:sn}
\Omega_{SN}=\frac{1}{2}\sum_{s=1}^S\sum_{m=1}^M\sum_{j\in \mathcal{F}_i} \mathcal{G}(\tau_{i,j}) (\textbf{w}^{s,m}\textbf{x}_i^{s,m}- \textbf{w}^{s,m}\textbf{x}_j^{s,m})^2,
\end{equation}
where $\mathcal{F}_i$ indicates a group contains students who are similar with student $i$, $\tau_{ij}$ is the similarity value between students $i$ and $j$, $\mathcal{G}(\tau_{i,j})$ is a function of $\tau_{i,j}$. Note that, since most co-occurrences are generated from students within the same major, we do not consider the social influence existing within students from different majors.

\begin{figure}[h]
	\centering
	\includegraphics[width=0.55\textwidth]{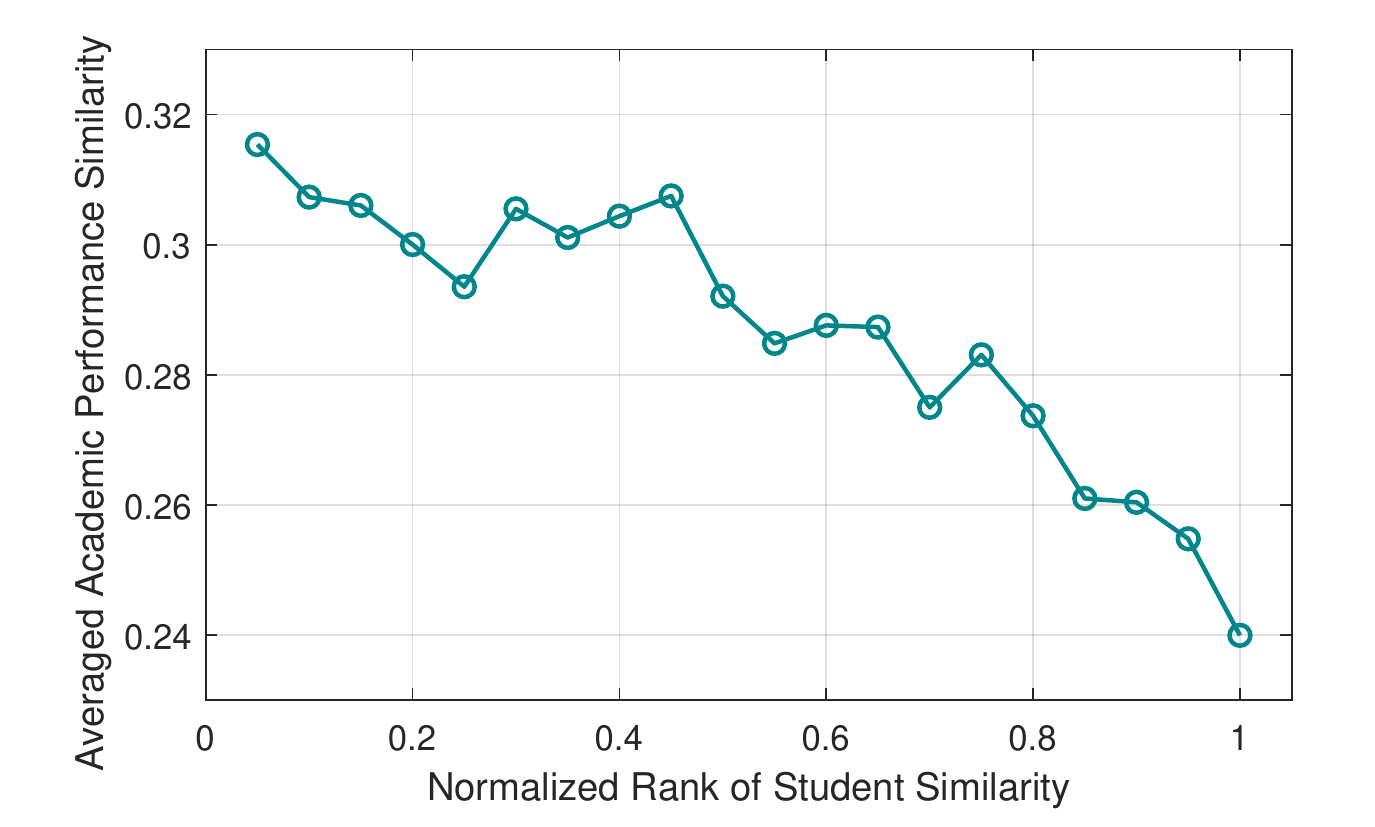}
	\caption{Correlation the strength of student similarity value and the similarity of academic performance}
	\label{fig:strength}
\end{figure}

In order to determine the formulation of function $\mathcal{G}$, we investigate the correlation between the strength of similarity and the influence of similar student. We rank the similarity value between students to 20 levels and then normalize it from $0.05-1$, and then calculate the average similarity of academic performance between students and their similar student in every similarity level. Its correlation with the strength of similarity is shown in Fig.~\ref{fig:strength}. The X-axis represents the average academic performance similarity and Y-axis represents the normalized rank of similarity value. The dots mean the average academic performance of each level of similarity value, and the blue line is a fitting curve. The result indicates that students' academic performance should be closer to those who are more similar with them.

The above analysis indicates that $\mathcal{G}$ should be an increasing function of $\tau_{i,j}$. When $\tau_{i,j}$ is larger, meaning student $i$ and $j$ are more similar, $\mathcal{G}_{\tau_{i,j}}$ should be larger. As shown in Figure~\ref{fig:strength}, in addition, the relation between the student similarity and the similarity of academic performance is close to linear. Thus, we define $\mathcal{G}(\tau_{i,j})=\tau_{i,j}$ in this work, and Eq.~\eqref{eq:sn} can be rewritten as:
\begin{equation}
\Omega_{SN}=\frac{1}{2}\sum_{s=1}^S\sum_{m=1}^M\sum_{j\in \mathcal{F}_i} \tau_{i,j} (\textbf{w}^{s,m}\textbf{x}_i^{s,m}- \textbf{w}^{s,m}\textbf{x}_j^{s,m})^2,
\end{equation}
Based on this regularization term, if students do not have enough data, we can infer their performance from those who are similar with them.

\subsection{Inference Task}
Without taking students' behavioral similarity into account, we can then predict the performance rank score for each student at the semester $s$ based on $f(\textbf{x}_i^{s,m}) = \textbf{u}^{s,m}\textbf{V}^s\textbf{x}^{s,m}_i$. Due to the academic performance similarity and the unreliability of data described before, it is necessary to integrate the performance score of each student with his/her similar students. In particular,
\begin{equation}\label{predict}
\hat{y}^{s,m}_i=(1-\xi)f(\textbf{x}_i^{s})+\xi \sum_{j\in \mathcal{F}_i}\frac{\tau_{ij}}{\sum_{k\in \mathcal{F}_i}\tau_{ik}}f(\textbf{x}_j^{s}),
\end{equation}
where the parameter $\xi$ balances the effect between behavioral factors and similar student's predicted performance. $\hat{y}^{s,m}_i$ is the final prediction score in this work. Then, we rank the final prediction score in each major of each semester and get the final rank list. 

\subsection{Optimization}
Based on the above discussion, by integrating inter-semester temporal correlation, inter-major correlation, and behavior similarity, we now formulate the whole loss function of multi-task learning-to-rank academic performance prediction (MTLTR-APP) framework as follows:
\begin{equation}\label{obj}
\begin{aligned}
\mathcal{L}(\textbf{U}^s,\textbf{V}^s)&=-\sum_{s=1}^{S}\sum_{m=1}^{M}\sum_{y_i^{s,m}>y_j^{s,m}}\log \sigma(\textbf{w}^{s,m}\textbf{x}_i^{s,m}-\textbf{w}^{s,m}\textbf{x}_j^{s,m})\\& +\frac{\lambda_{s}}{2}\sum_{s=1}^{S-1}\Vert \textbf{W}^{s} - \textbf{W}^{s+1} \Vert_F^2+\sum_{s=1}^{S}\lambda_1\Vert \textbf{U}^s\Vert_1
+\lambda_2\sum_{s=1}^{S}\Vert \textbf{V}^s\Vert_2^2\\&+\frac{\lambda_{n}}{2}\sum_{s=1}^S\sum_{m=1}^M\sum_{j\in \mathcal{F}_i} \tau_{i,j} (\textbf{w}^{s,m}\textbf{x}_i^{s,m}- \textbf{w}^{s,m}\textbf{x}_j^{s,m})^2\\
&s.t., \textbf{W}^s=\textbf{U}^s\textbf{V}^s,\;\textbf{U}^s\succeq \textbf{0},
\end{aligned}
\end{equation}
where $\lambda_{s}$, $\lambda_{n}$ are used to control the effect of inter-semester temporal correlation and behavior similarity, respectively. The greater the value of each $\lambda$ (i.e., $\lambda_{n}$, $\lambda_{s}$, $\lambda_1$, $\lambda_2$), the stronger effect of its corresponded part. 

Then, we optimize the constrained loss function using a block coordinate gradient descent algorithm, which solving $\textbf{U}^{s}$ and $\textbf{V}^{s}$ alternately. In addition, we also solve each $\textbf{u}^{s,m}$ alternatively. We describe the details as follows:
\subsubsection{Fix every $\textbf{U}^{s}$, solve $\textbf{V}^s$}
When we fix $\textbf{U}^{s}$, the loss function becomes:
\begin{equation}
\label{eq:gradv}
\begin{aligned}
\mathcal{L}(\textbf{V}^{s})&=-\sum_{s=1}^{S}\sum_{m=1}^{M}\sum_{y_i^{s,m}>y_j^{s,m}}\log\sigma(\textbf{u}^{s,m}\textbf{V}^{s}\textbf{x}_i^{s,m}-\textbf{u}^{s,m}\textbf{V}^{s}\textbf{x}_j^{s,m})\\& +\frac{\lambda_{s}}{2}\sum_{s=1}^{S-1}\Vert \textbf{U}^{s}\textbf{V}^{s} - \textbf{U}^{s+1}\textbf{V}^{s+1} \Vert_F^2+\lambda_2\sum_{s=1}^S\Vert \textbf{V}^s\Vert_2^2\\&
+\frac{\lambda_{n}}{2}\sum_{s=1}^S\sum_{m=1}^M\sum_{j\in \mathcal{F}_i} \tau_{i,j}(\textbf{u}^{s,m}\textbf{V}^{s}\textbf{x}_i^{s,m}- \textbf{u}^{s,m}\textbf{V}^{s}\textbf{x}_j^{s,m})^2.
\end{aligned}
\end{equation}
It can be solved by stochastic gradient descent (SGD). The gradient of Eq.~\eqref{eq:gradv} is . 
\begin{equation}
\label{eq:optimv}
	\begin{aligned}
		&\nabla\mathcal{L}(\textbf{V}^s) =\sum_{m=1}^{M}\sum_{y_i^{s,m}>y_j^{s,m}}( \sigma(\textbf{u}^{s,m}\textbf{V}^{s}\textbf{x}_i^{s,m}-\textbf{u}^{s,m}\textbf{V}^{s}\textbf{x}_j^{s,m})-1)(\textbf{u}^{s,m})^T\\&(\textbf{x}_i^{s}-\textbf{x}_j^{s})^T +\lambda_{s}(\textbf{U}^s)^{T}(2\textbf{U}^s\textbf{V}^s-\textbf{U}^{s+1}\textbf{V}^{s+1}-\textbf{U}^{s-1}\textbf{V}^{s-1})+2\lambda_2\textbf{V}^s\\&
		+\lambda_{n}\sum_{m=1}^M\sum_{j\in \mathcal{F}_i} \tau_{i,j} (\textbf{u}^{s,m}\textbf{V}^{s}\textbf{x}_i^{s}- \textbf{u}^{s,m}\textbf{V}^{s}\textbf{x}_j^{s,m})(\textbf{u}^{s,m})^T(\textbf{x}_i^{s}-\textbf{x}_j^{s,m})^T.
	\end{aligned}
\end{equation}
%
\subsubsection{Fix $\textbf{V}^s$, solve $\textbf{u}^{s,m}$} 
When we fix $\textbf{V}^s$, the loss function of becomes:
\begin{equation}
\label{eq:gradu}
\begin{aligned}
\mathcal{L}(\textbf{u}^{s,m})&=-\sum_{s=1}^{S}\sum_{m=1}^{M}\sum_{y_i^{s,m}>y_j^{s,m}}\log \sigma(\textbf{u}^{s,m}\textbf{V}^{s}\textbf{x}_i^{s,m}-\textbf{u}^{s,m}\textbf{V}^{s}\textbf{x}_j^{s,m})\\& +\frac{\lambda_{s}}{2}\sum_{s=1}^{S-1}\sum_{m=1}^{M}\Vert \textbf{u}^{s,m}\textbf{V}^{s} - \textbf{u}^{s+1,m}\textbf{V}^{s+1} \Vert_F^2+\lambda_1\sum_{s=1}^{S}\sum_{m=1}^{M}\Vert \textbf{u}^{s,m}\Vert_1\\&
+\frac{\lambda_{n}}{2}\sum_{s=1}^S\sum_{m=1}^M\sum_{j\in \mathcal{F}_i} \tau_{i,j} (\textbf{u}^{s,m}\textbf{V}^{s}\textbf{x}_i^{s,m}- \textbf{u}^{s,m}\textbf{V}^{s}\textbf{x}_j^{s,m})^2\\
&s.t., \textbf{U}^s\succeq \textbf{0}.
\end{aligned}
\end{equation}
This optimization problem can be efficiently solved by using proximal gradient descent, which is a common used optimization method to solve optimization problem with non-differentiable part. Let $\mathcal{L}(\textbf{u}^{s,m})=\mathcal{L}_g(\textbf{u}^{s,m})+\mathcal{L}_h(\textbf{u}^{s,m})$, where $\mathcal{L}_h(\textbf{u}^{s,m})=\lambda_1\sum_{s=1}^{S}\sum_{m=1}^{M}\Vert \textbf{u}^{s,m}\Vert_1$, $\mathcal{L}_g$ is the rest part of Eq.~\eqref{eq:gradu}. Then, the gradient of $\mathcal{L}_g(\textbf{u}^{s,m})$ is:
\begin{equation}
\label{eq:optimu}
	\begin{aligned}
	&\nabla\mathcal{L}_g(\textbf{u}^{s,m}) =\sum_{y_i^{s,m}>y_j^{s,m}}(\sigma(\textbf{u}^{s,m}\textbf{V}^s\textbf{x}_i^{s,m}-\textbf{u}^{s,m}\textbf{V}^s\textbf{x}_j^{s,m})-1)\\&(\textbf{x}_i^{s,m}-\textbf{x}_j^{s,m})^T(\textbf{V}^s)^T +\lambda_{s}(2\textbf{u}^{s,m}\textbf{V}^s - \textbf{u}^{s+1,m}\textbf{V}^{s+1}-\textbf{u}^{s-1}\textbf{V}^{s-1})\\&(\textbf{V}^s)^T
		+\lambda_{n}\sum_{j\in \mathcal{F}_i} \tau_{i,j} (\textbf{u}^{s,m}\textbf{V}^{s}\textbf{x}_i^{s}- \textbf{u}^{s,m}\textbf{V}^{s}\textbf{x}_j^{s})(\textbf{x}_i^{s,m}-\textbf{x}_j^{s,m})^T(\textbf{V}^s)^T.\\
	\end{aligned}
\end{equation}
The proximal mapping is defined as $\mathrm{prox}_t(x)=\arg\min_z\frac{1}{2t}\Vert x-z\Vert_2^2+\lambda\Vert z\Vert_1$. The solution of it is:
\begin{equation}
\centering
z_i=\left\{  
\begin{array}{lr}  
x_i-\lambda t &  x_i>\lambda t,\\  
0, & -\lambda t\leq x_i\leq \lambda t,\\  
x_i+\lambda t & x_i<-\lambda t .  
\end{array}  
\right.
\end{equation}
In our optimization problem. $\mathrm{prox}_t(x)=\mathrm{prox}_t(\textbf{U}^s-t\nabla\mathcal{L}_g(\textbf{U}^{s}))$. A further projection is used to ensure the elements of $\textbf{U}^{s}$ are non-negative, which is defined as $\mathcal{B}(x)=max(x,0)$. Furthermore, The whole optimization framework is shown in Algorithm~\ref{alg:app}.
\begin{algorithm}[h]
	\caption{Optimization Framework of MTLTR-APP}
	\label{alg:app}
	\KwIn{Given feature $\{\textbf{X}^{1,1}, ..., \textbf{X}^{s,m}, ..., \textbf{X}^{S,M}\}$, academic performance $\{\textbf{y}^{1,1}, ..., \textbf{y}^{s,m}, ..., \textbf{y}^{S,M}\}$, coefficient parameters $\lambda_{s}$, $\lambda_{n}$, $\lambda_{1}$, $\lambda_{2}$, $\xi$}
	\KwOut{Optimized $\textbf{U}^s$, $\textbf{V}^s$}
	\BlankLine
	random initialize $\textbf{U}^s$, $\textbf{V}^s$\;
	\While{not convergence}{
		\For{semester $s$ in $\{1, ..., S\}$}{
		Optimize $\textbf{V}^s$ based on Eq.~\eqref{eq:optimv}\;
		\For{major $m$ in $\{1,...,M\}$}{
			Calculate gradient $\nabla \mathcal{L}_g(\textbf{u}^{s,m})$ based on Eq.~\eqref{eq:optimu}\;
			Update $\textbf{u}^{s,m}$ by $\mathrm{prox}_t(\textbf{u}^{s,m}-t\nabla\mathcal{L}_g(\textbf{u}^{s,m}))$\;
			Use projection $\mathcal{B(\cdot)}$ to make $\textbf{u}^{s,m}$ non-negative\;
			}
		}
	}
\end{algorithm}

The time complexity of one gradient descent iteration in our optimization algorithm is $O(N^2SMpk)$. $N$, $M$, $S$, $p$, $k$ is the maximum number of students among all majors, the number of majors, the number of semesters, the length of feature vector, the number of hidden categories, respectively. $M$ and $S$ is often less than 50 and 10, respectively. 

\section{Experiment}
\subsection{Experimental Settings and Data Description}
The correlation between behavioral factors and academic performance varies from semester to semester. Thus, for evaluating the proposed algorithm, we train our algorithm on a grade of $3,352$ college students from $18$ majors and test it on the subsequent grade of $3,245$ students from $17$ majors. Both grades data include the records in the first $5$ semesters. Furthermore, we randomly select 10\% data of training set as validation set. All hyperparameters are tuned based on the accuracy in validation set. Specifically, in our experiment, we set the size of hidden variable as 5 (i.e. $k=5$). We set $\lambda_s$, $\lambda_n$, $\lambda_1$, $\lambda_2$, $\xi$ as $1$, $0.01$, $0.5$, $0.1$, $0.2$, respectively.

In addition, as we described before, the collection of the behavioral data benefits from the existing information management and operation systems. Currently, these behavioral data include students' consumption history at different locations (e.g., cafeterias, campus supermarket), entry-exit records in the library and dormitory, and the history of borrowing books. Besides, these data are accompanied by students' GPA information from each semester. For protecting privacy, we not only anonymize students' sensitive information and randomly assign each student a unique code, but also convert the GPA information into the relative performance ranking of any two students. The data statistics are illustrated in Table~\ref{tab:dataset}. 
\begin{table}[h]
	\centering
	\caption{Dataset Statistics}
	\begin{tabular}{l|c|c}
		\hline
		& Training & Testing\\\hline
		Num. of Student & 3,352 & 3,245\\
		Num. of Major & 18 & 17\\
		Num. of semesters & 5 & 5\\\hline
		Num. of consumption & 10,390,715 & 9,866,884\\
		Num. of library entrance \& exit & 781,630 & 727,854\\
		Num. of borrowing history & 218,719 & 215,413\\
		Num. of Grades & 126,683 & 124,291\\\hline
	\end{tabular}
	\label{tab:dataset}
\end{table}

\subsection{Evaluation Metric}
Since we convert students' GPA into the performance rank out of privacy concern, we assess the performance of the proposed algorithm with ranking-based metrics. Assume the rank prediction of all students are equally important, we exploit the \emph{Spearman rank correlation} for measuring the correlation between the predicted rank and the actual rank. The higher the Spearman coefficient, the better the prediction performance. When testing, we are concerned with the prediction performance of the proposed algorithm at each semester $t$, which is defined as, after averaging over majors:
\begin{equation}
\rho_t=\frac{6\sum_{i=1}^{|\mathcal{N}|}(a_i^t-b_i^t)}{|\mathcal{N}|(|\mathcal{N}|^2-1)},
\end{equation}
where $|\mathcal{N}|$ is the number of students. And $a_i^t$ means the real ranking of student $i$ in semester $t$, while $b_i^t$ is predicted ranking.
\subsection{Compared Methods}
Based on the evaluation protocol, we compare the proposed algorithm with the following competing baselines. All methods use the features listed in Table~\ref{tab:feature}. 
\begin{itemize}
	\item \textbf{Ridge Regression}: Similar as SmartGPA~\cite{rui2015smartgpa}, we use ridge regression (i.e., original linear regression with $\ell_2$-norm regularization) for academic performance prediction.
	\item \textbf{Decision Tree (DT)}:  We train a decision tree for academic performance prediction.
	\item \textbf{Random Forest (RF)}: The random forest is used for academic performance prediction.
	\item \textbf{XGBoost (XGB)~\cite{chen2016xgboost}}: XGBoost is a boosting-tree-based method and is widely used in various data mining applications.
	\item \textbf{Multiple layer perceptron (MLP)}: Our method is compared with neural network of four fully connected layers. The number of hidden unites are 64, 128, 128, and 64, respectively.
	\item \textbf{RankSVM~\cite{joachims2009svm}}: We compare our method with RankSVM for academic performance prediction, which is a pair-wise learning-to-rank model based on SVM.
\end{itemize}
Note that, in decision tree, random forest, and multiple layer perceptron, XGBoost, we also convert students' academic performance to their relative performance, and use these baselines classify the relative performance.

We also study the effect of different variants of our method. The corresponded loss function are also listed. The term $\lambda_e\sum_{s=1}^S\Vert \mathbf{W}^s\Vert_2^2$ is used for the regularization.
\begin{itemize}
	\item \textbf{BLTR}: Basic learning-to-rank model (loss defined in Eq.~\eqref{eq:ranknet}), considering the academic performance ranking as a whole task. In other words, weight is constant with the semesters. The loss function is:
	\begin{equation}\label{obj_bltr}
    \mathcal{L}(\textbf{W}^s)=-\sum_{s=1}^{S}\sum_{m=1}^{M}\sum_{y_i^{s,m}>y_j^{s,m}}\log \sigma(\textbf{w}^{s,m}\textbf{x}_i^{s,m}-\textbf{w}^{s,m}\textbf{x}_j^{s,m}) + \lambda_e\sum_{s=1}^S\Vert \mathbf{W}^s\Vert_2^2
    \end{equation}
	\item \textbf{BLTR+SS}: Imposing the basic learning-to-rank model with student similarity incorporated. The loss function is:
	\begin{equation}\label{obj}
    \begin{aligned}
    \mathcal{L}(\textbf{W}^s)&=-\sum_{s=1}^{S}\sum_{m=1}^{M}\sum_{y_i^{s,m}>y_j^{s,m}}\log \sigma(\textbf{w}^{s,m}\textbf{x}_i^{s,m}-\textbf{w}^{s,m}\textbf{x}_j^{s,m})+ \lambda_e\sum_{s=1}^S\Vert \mathbf{W}^s\Vert_2^2\\&+\frac{\lambda_{n}}{2}\sum_{s=1}^S\sum_{m=1}^M\sum_{j\in \mathcal{F}_i} \tau_{i,j} (\textbf{w}^{s,m}\textbf{x}_i^{s,m}- \textbf{w}^{s,m}\textbf{x}_j^{s,m})^2\\
    \end{aligned}
    \end{equation}
	\item \textbf{BLTR+MS}: Major-based BLTR, with integrating inter-major correlation. The loss function is:
	\begin{equation}\label{obj}
    \begin{aligned}
    \mathcal{L}(\textbf{U}^s,\textbf{V}^s)&=-\sum_{s=1}^{S}\sum_{m=1}^{M}\sum_{y_i^{s,m}>y_j^{s,m}}\log \sigma(\textbf{w}^{s,m}\textbf{x}_i^{s,m}-\textbf{w}^{s,m}\textbf{x}_j^{s,m})+\sum_{s=1}^{S}\lambda_1\Vert \textbf{U}^s\Vert_1
    +\lambda_2\sum_{s=1}^{S}\Vert \textbf{V}^s\Vert_2^2\\&+\frac{\lambda_{n}}{2}\sum_{s=1}^S\sum_{m=1}^M\sum_{j\in \mathcal{F}_i} \tau_{i,j} (\textbf{w}^{s,m}\textbf{x}_i^{s,m}- \textbf{w}^{s,m}\textbf{x}_j^{s,m})^2\\
    &s.t., \textbf{W}^s=\textbf{U}^s\textbf{V}^s,\;\textbf{U}^s\succeq \textbf{0},
    \end{aligned}
    \end{equation}
	\item \textbf{BLTR+SEQ}: Semester-based BLTR, which combines BLTR with sequential smoothness regularization. The loss function is:
	\begin{equation}\label{obj_bltrseq}
    \mathcal{L}(\textbf{W}^s)=-\sum_{s=1}^{S}\sum_{m=1}^{M}\sum_{y_i^{s,m}>y_j^{s,m}}\log \sigma(\textbf{w}^{s,m}\textbf{x}_i^{s,m}-\textbf{w}^{s,m}\textbf{x}_j^{s,m})+\frac{\lambda_{s}}{2}\sum_{s=1}^{S-1}\Vert \textbf{W}^{s} - \textbf{W}^{s+1} \Vert_F^2+ \lambda_e\sum_{s=1}^S\Vert \mathbf{W}^s\Vert_2^2 
    \end{equation}
	\item \textbf{MTLTR-APP}: The proposed model, imposing sequential smoothness regularization, inter-major correlation on BLTR, incorporating student similarity.
\end{itemize}
\subsection{Performance Comparison}
\subsubsection{Comparison with baseline methods}
We test our method and report the average Spearman correlation of each major in each semester. The results are shown in Table~\ref{tab:answer}. By comparing MTLTR-APP with the best baseline, MTLTR-APP achieves the highest Spearman Correlation in every semester. More specifically, we can see ridge regression performs worst, as it only regards the ranking problem as regression. It is difficult to predict the rank directly since the rank list is converted from GPA. The distribution of GPA is usually Gaussian distribution, while rank is equally spaced. DT, RF, RankSVM, and MLP, XGB also use the pair-wise method to convert the ranking problem as the binary classification problem. It significantly performs better than Ridge. However, these methods do not model the inter-semester and inter-major correlation. In addition, it still overlooks the student similarity. Our proposed MTLTR-APP integrate these informations and achieves the best performance.

\begin{table}[h]
	\centering
	\caption{Comparison of MTLTR-APP with baselines}
	\begin{tabular}{l|c|c|c|c|c}
		\hline
		Semester & 1 & 2 & 3 & 4 & 5\\\hline
		Ridge Regression & 0.235 & 0.365 & 0.455 & 0.453 & 0.477\\
		Decision Tree & 0.267 & 0.366 & 0.460 & 0.460 & 0.489\\
		Random Forest & 0.278 & 0.377 & 0.463 & 0.468 & 0.495\\
		XGBoost & 0.281 & 0.379 & 0.465 & 0.470 & 0.497\\
		RankSVM & 0.275 & 0.374 & 0.461 & 0.461 & 0.496\\
		MLP & 0.280 & 0.375 & 0.466 & 0.465 & 0.498\\\hline
		MTLTR-APP & \textbf{0.300} & \textbf{0.389} & \textbf{0.481} & \textbf{0.484} & \textbf{0.513} \\
		\hline
	\end{tabular}
	\label{tab:answer}
\end{table}
\subsubsection{Comparison with variants of our proposed method}
Table~\ref{tab:variant} shows the performance of MTLTR-APP and its variants. We also test our method and report the average Spearman correlation of each major in each semester. By comparing BLTR+SS with BLTR, \emph{first}, we can see that student similarity does make the improvement to academic performance prediction. \emph{Second}, by comparing BLTR+MS with BLTR, due to the better performance of BLTR+MS, we conclude the necessity of inter-major correlation. This also implies the inter-major correlation. \emph{Third}, by comparing BLTR+SEQ with BLTR, the superiority of the former to the latter indicates the benefit of applying sequential smoothness regularization. This also implies the varying correlation across different semesters. \emph{Finally}, the proposed algorithm, MTLTR-APP, can achieve the better performance, compared to each variant, after the incorporation of inter-semester correlation, inter-major correlation and student similarity, which shows the effectiveness of the integrated model.

\begin{table}[h]
	\centering
	\caption{Comparison of MTLTR-APP with its variants}
	\begin{tabular}{l|c|c|c|c|c}
		\hline
		Semester & 1 & 2 & 3 & 4 & 5\\\hline
		BLTR & 0.273 & 0.372 & 0.453 & 0.461 & 0.493 \\
		BLTR+SS & 0.286 & 0.385 & 0.464 & 0.466 & 0.497\\
		BLTR+MS & 0.289 & 0.383 & 0.471 & 0.473 & 0.502 \\
		BLTR+SEQ & 0.294 & 0.376 & 0.466 & 0.475 & 0.507 \\
		MTLTR-APP & \textbf{0.300} & \textbf{0.389} & \textbf{0.481} & \textbf{0.484} & \textbf{0.513} \\
		
		\hline
	\end{tabular}
	\label{tab:variant}
\end{table}
\subsection{Feature Importance}
In order to compare the importance of different types of features (diligence, orderliness and sleep patterns) for academic performance prediction, we evaluate the prediction results under various feature combinations. In this experiment, we run our proposed model MTLTR-APP on five consecutive semesters, semester 1-5. Since the features of sleep pattern are 0-1 feature vectors which transferred by one hot, so we do not use it alone to predict the academic performance. The results of the Spearman correlation under different settings of feature combinations are shown in Table~\ref{tab:feature_comp}.

In this table, \emph{first}, we can see that using diligence or orderliness alone to predict academic performance can get the similar results (see column 1 vs. column 2), and adding them together can improve the accuracy significantly (see column 3 vs. column 1 and 2). Thus, the diligence and orderliness can reflect students' study in different ways, which is consistent with our motivation before: diligence reflects how much time students spend on the study while orderliness depicts the regularity of students' lifestyle. \emph{Second}, adding sleep patterns only improve slightly (see column 4 vs. column 1, column 5 vs. column 2, column 6 vs. column 3). One potential reason is that we do not have the actual information about students' sleep pattern, we only use the first and last time of smartcard records to represent the wake-up time and bedtime, which may be insufficient. In conclusion, each feature type extracted from different dimensions of students' lifestyle is important for academic performance prediction.

\begin{table}[h]
	\centering
	\caption{Performance evaluation under Different Feature Conditions.}
	\label{tab:feature_comp}
	\begin{tabular}{|c|c|c|c|c|c|c|c|}
		\hline
		\multicolumn{2}{|c|}{}          & \multicolumn{6}{c|}{Settings}                                                                                                           \\ \hline
		\multicolumn{2}{|c|}{Column ID} & 1                    & 2                    & 3                    & 4                    & 5                    & 6                    \\ \hline
		\multirow{3}{*}{Features$^1$}  & D  & \checkmark &                      & \checkmark & \checkmark &                      & \checkmark \\ \cline{2-8} 
		& O  &                      & \checkmark & \checkmark &                      & \checkmark & \checkmark \\ \cline{2-8} 
		& SP &                      &                      &                      & \checkmark & \checkmark & \checkmark \\ \hline
		\multicolumn{2}{|c|}{Semester}  & \multicolumn{6}{c|}{Spearman Correlation}                                                                                               \\ \hline
		\multicolumn{2}{|c|}{1}         & 0.230                & 0.249                & 0.298                & 0.231                & 0.251                & \textbf{0.300}                \\ \hline
		\multicolumn{2}{|c|}{2}         & 0.338                & 0.352                & 0.385                & 0.338                & 0.352                & \textbf{0.389}                \\ \hline
		\multicolumn{2}{|c|}{3}         & 0.428                & 0.438                & 0.477                & 0.431                & 0.439                & \textbf{0.481}                \\ \hline
		\multicolumn{2}{|c|}{4}         & 0.436                & 0.436                & 0.481                & 0.438                & 0.435                & \textbf{0.484}                \\ \hline
		\multicolumn{2}{|c|}{5}         & 0.441                & 0.455                & 0.511                & 0.445                & 0.460                & \textbf{0.513}                \\ \hline
	\end{tabular}
	\vspace{0.1em}\\
	\footnotesize{$^1$ D, O, SP mean Diligence, Orderliness, Sleep Patterns, respectively\\}
\end{table}

\subsection{Sensitivity Analysis}
\begin{figure}[h]
	\centering
	\subfigure[Effect of $\lambda_m$]{
		\centering
		\label{fig:para_mtrnt}
		\includegraphics[width=0.35\textwidth]{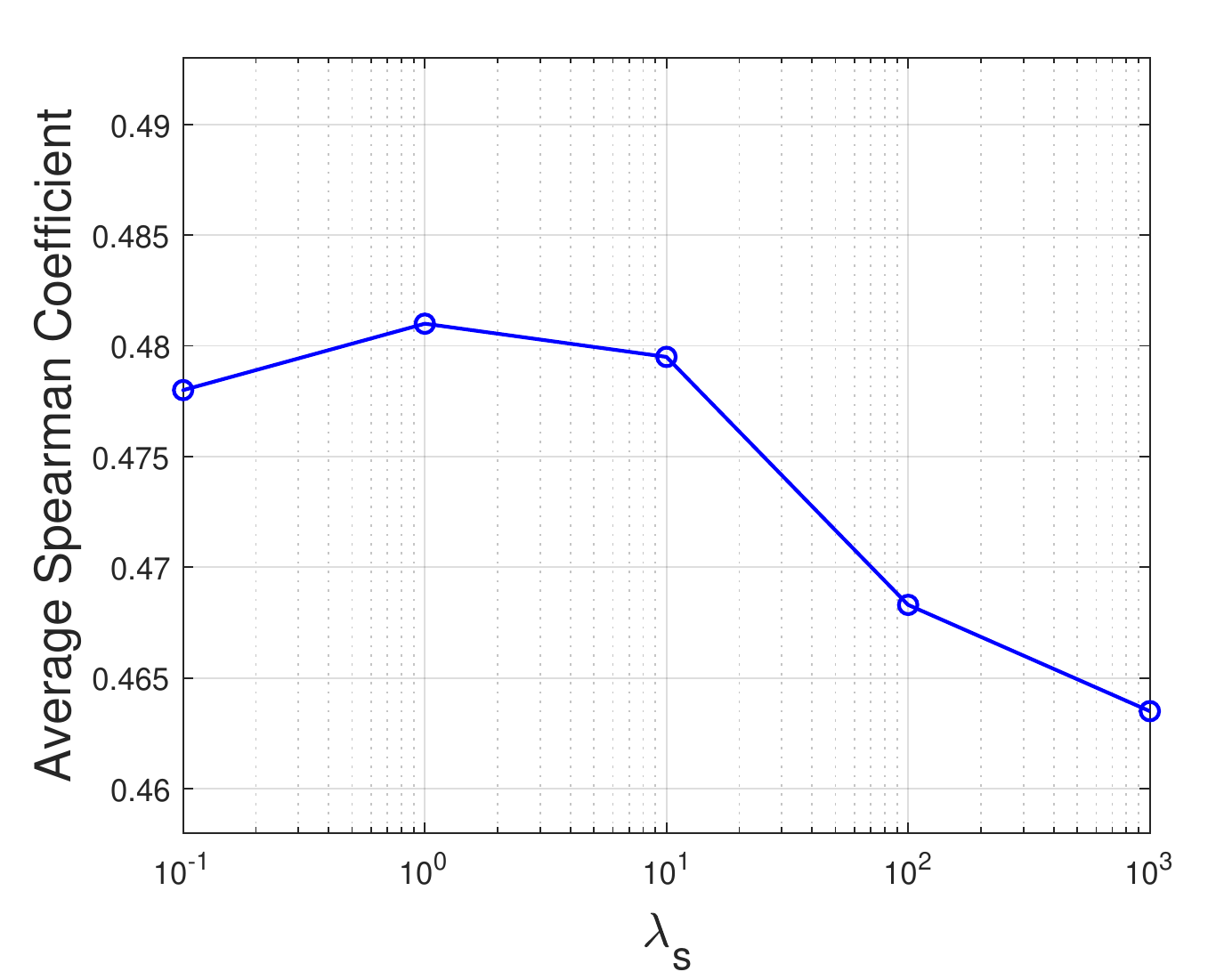}}
	\subfigure[Effect of $\xi$]{
		\centering
		\label{fig:para_test}
		\includegraphics[width=0.35\textwidth]{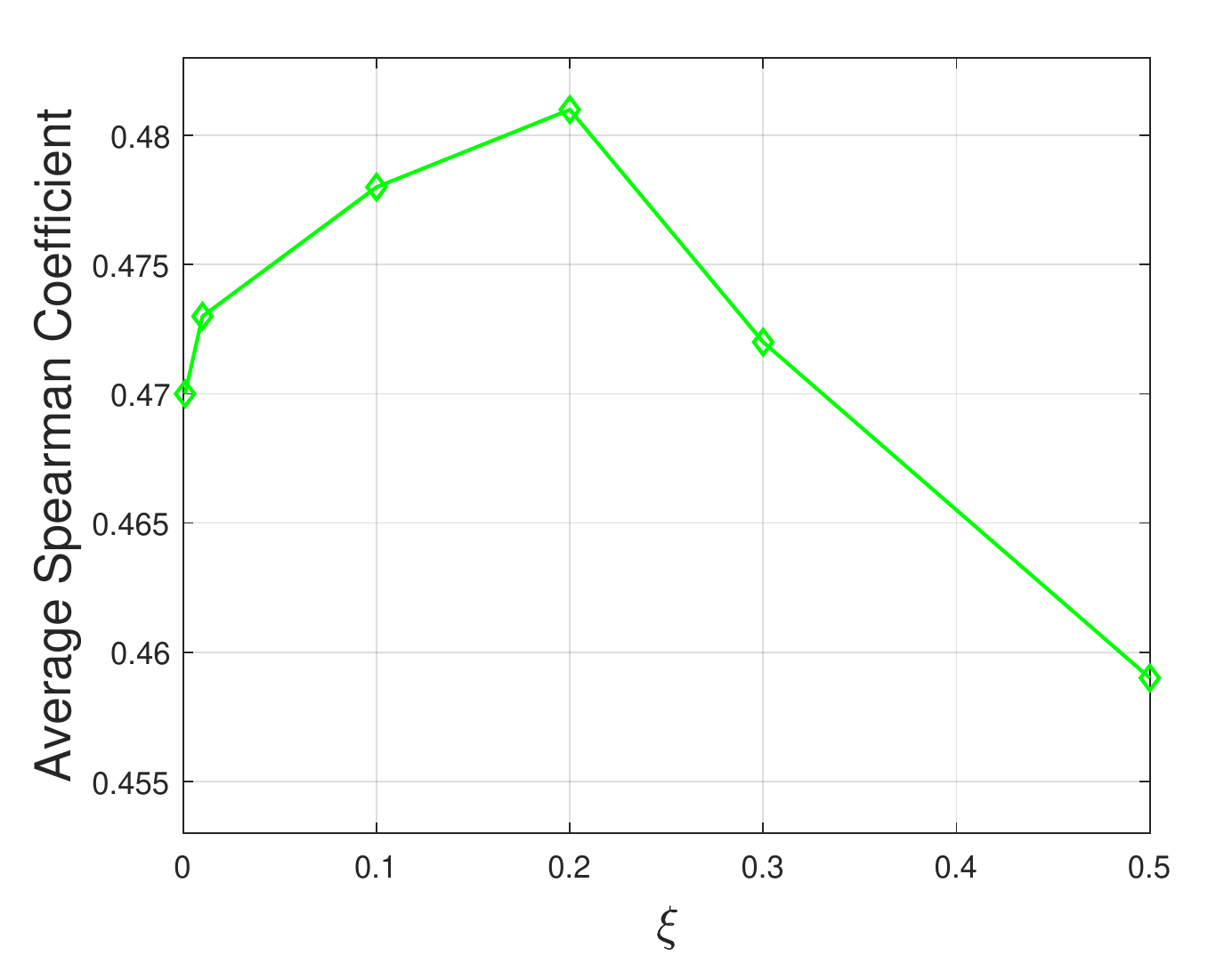}}
	\caption{(a) The effect of sequentially smoothness. (b) The effect of the academic performance similarity.}
	\label{statistical}
\end{figure}

Although we have observed the effectiveness of MTLTR-APP, how the performance changes with the regularization coefficient is still under exploration. In this studies, we investigate the effect of sequentially smoothness and academic performance similarity. The results of these studies in semester 3 are shown in Fig~\ref{statistical}.  We can see that the relatively optimal values for the $\lambda_s$ are $1.0$. This indicates that the sequentially smoothness regularization play an important part in improving learning the weight of each task. Besides, the optimal value of the parameter $\xi$ in Eq.~\eqref{predict} is set around $0.2$, indicating incorporating student similarity could improve academic performance prediction.

\section{Conclusion and Discussion}
In this paper, we analyze a large-scale students' behavioral data on campus for predicting academic performance. We find that there are significant correlations between students characteristics -- diligence, orderliness and sleep pattern -- with academic performance. 
Besides, frequently co-occurring students have close academic performance. Then we design a multi-task learning-to-rank academic performance prediction framework (MTLTR-APP), which captures the inter-major and inter-semester dependency and integrate student similarity. By training the proposed algorithm on a grade of students and testing it on the subsequent grade of students, we show the effectiveness of our proposed MTLTR-APP for predicting academic performance and demonstrate the inter-semester correlation, inter-major correlation, and student similarity are effective in academic performance prediction. Finally, we show the importance of each characteristic for academic performance prediction. Our findings and proposed framework can be further extended in the following directions.

In this paper, due to the limitation of data collection, some important factors can not be captured and used in our framework, such as intellectual level, psychology factors, sport participation, and even the luck in the exam. In addition, we can not capture all the daily behavior of students. For example, some students prefer study in their dormitories rather than teaching building. Some students show abnormal behavior may achieve great performance. We will collect more relevant data in future research.

In addition, in-class performance is very important for academic performance prediction. In the future, we will collected more in-class behavioral data and incorporate it with offline campus daily behavior to enhance the prediction model. We also plan to integrate the academic performance framework to modern educational management system. Then, we are able to use real-time data to predict students' academic performance. The educators can guide students and give them suggestions in advance based on the prediction results.

\begin{acks}

This work was partially supported by the National Natural Science Foundation of China (61603074, 61473060, 61433014, 61502083). Tao Zhou is supported by the Science Promotion Programme of UESTC (no. Y03111023901014006). Defu Lian is supported by the National Natural Science Foundation of China (Grant No. 61502077 and 61631005) and the Fundamental Research Funds for the Central Universities (Grant No. ZYGX2016J087).

\end{acks}

\bibliographystyle{ACM-Reference-Format}
\bibliography{ref}

\end{document}